\documentclass[manyauthors,nocleardouble,COMPASS]{cernphprep}
\usepackage{graphicx}
\usepackage{rotating}
\usepackage{subfig}
\usepackage[percent]{overpic}
\usepackage[numbers, square, comma, sort&compress]{natbib}

\usepackage{hyperref}

\newcommand{\leftexp}[2]{{\vphantom{#2}}^{#1}{#2}}

\newcommand{\mrf}[1]{\mbox{$\mathrm{#1}$}}
\newcommand{\mif}[1]{\mbox{$\mathit{#1}$}}

\newcommand{\gom}{\mrf{GeV/\mif{c}}}

\newcommand{\gomt}{\mrf{(GeV/\mif{c})^2}}

\begin{document}

\begin{titlepage}
\PHnumber{2013--91}
\PHdate{May 30, 2013}

\title{Hadron Transverse Momentum Distributions in Muon \\ Deep Inelastic Scattering at 160 GeV/$c$}

\Collaboration{The COMPASS Collaboration}
\ShortAuthor{The COMPASS Collaboration}

\begin{abstract}
\label{sec:abstract}
Multiplicities of charged hadrons produced in deep inelastic muon scattering off
a $^6$LiD target have been measured as a function of the DIS variables $x_{Bj}$,
$Q^2$, $W^2$ and the final state hadron variables $p_T$ and $z$. The $p_T^2$
distributions are fitted with a single exponential function at low values of
$p_T^2$ to determine the dependence of $\langle p_T^2 \rangle$ on $x_{Bj}$,
$Q^2$, $W^2$ and $z$. The $z$-dependence of $\langle p_T^2 \rangle$ is shown to
be a potential tool to extract the average intrinsic transverse momentum squared
of partons, $\langle k_{\perp}^2 \rangle$, as a function of $x_{Bj}$ and $Q^2$
in a leading order QCD parton model.
\end{abstract}

\vfill
\Submitted{(to be submitted to European Physical Journal C)}
\end{titlepage}

{\pagestyle{empty}
%
%

\section*{The COMPASS Collaboration}
\label{app:collab}
\renewcommand\labelenumi{\textsuperscript{\theenumi}~}
\renewcommand\theenumi{\arabic{enumi}}
\begin{flushleft}
C.~Adolph\Irefn{erlangen},
M.G.~Alekseev\Irefn{triest_i},
V.Yu.~Alexakhin\Irefn{dubna},
Yu.~Alexandrov\Irefn{moscowlpi}\Deceased,
G.D.~Alexeev\Irefn{dubna},
A.~Amoroso\Irefn{turin_u},
V.~Andrieux\Irefn{saclay},
A.~Austregesilo\Irefnn{cern}{munichtu},
B.~Bade{\l}ek\Irefn{warsawu},
F.~Balestra\Irefn{turin_u}
% ...
\end{flushleft}

%
%

\begin{Authlist}
\item \Idef{bielefeld}{Universit\"at Bielefeld, Fakult\"at f\"ur Physik, 33501 Bielefeld, Germany\Arefs{f}}
\item \Idef{bochum}{Universit\"at Bochum, Institut f\"ur Experimentalphysik, 44780 Bochum, Germany\Arefs{f}}
\item \Idef{bonniskp}{Universit\"at Bonn, Helmholtz-Institut f\"ur  Strahlen- und Kernphysik, 53115 Bonn, Germany\Arefs{f}}
\item \Idef{bonnpi}{Universit\"at Bonn, Physikalisches Institut, 53115 Bonn, Germany\Arefs{f}}
\item \Idef{brno}{Institute of Scientific Instruments, AS CR, 61264 Brno, Czech Republic\Arefs{g}}
\item \Idef{calcutta}{Matrivani Institute of Experimental Research \& Education, Calcutta-700 030, India\Arefs{h}}
\item \Idef{dubna}{Joint Institute for Nuclear Research, 141980 Dubna, Moscow region, Russia\Arefs{i}}
\item \Idef{erlangen}{Universit\"at Erlangen--N\"urnberg, Physikalisches Institut, 91054 Erlangen, Germany\Arefs{f}}
\item \Idef{freiburg}{Universit\"at Freiburg, Physikalisches Institut, 79104 Freiburg, Germany\Arefs{f}}
\item \Idef{cern}{CERN, 1211 Geneva 23, Switzerland}
\item \Idef{liberec}{Technical University in Liberec, 46117 Liberec, Czech Republic\Arefs{g}}
\item \Idef{lisbon}{LIP, 1000-149 Lisbon, Portugal\Arefs{j}}
\item \Idef{mainz}{Universit\"at Mainz, Institut f\"ur Kernphysik, 55099 Mainz, Germany\Arefs{f}}
\item \Idef{miyazaki}{University of Miyazaki, Miyazaki 889-2192, Japan\Arefs{k}}
\item \Idef{moscowlpi}{Lebedev Physical Institute, 119991 Moscow, Russia}
\item \Idef{munichlmu}{Ludwig-Maximilians-Universit\"at M\"unchen, Department f\"ur Physik, 80799 Munich, Germany\Arefs{f}\Arefs{l}}
\item \Idef{munichtu}{Technische Universit\"at M\"unchen, Physik Department, 85748 Garching, Germany\Arefs{f}\Arefs{l}}
\item \Idef{nagoya}{Nagoya University, 464 Nagoya, Japan\Arefs{k}}
\item \Idef{praguecu}{Charles University in Prague, Faculty of Mathematics and Physics, 18000 Prague, Czech Republic\Arefs{g}}
\item \Idef{praguectu}{Czech Technical University in Prague, 16636 Prague, Czech Republic\Arefs{g}}
\item \Idef{protvino}{State Research Center of the Russian Federation, Institute for High Energy Physics, 142281 Protvino, Russia}
\item \Idef{saclay}{CEA IRFU/SPhN Saclay, 91191 Gif-sur-Yvette, France}
\item \Idef{telaviv}{Tel Aviv University, School of Physics and Astronomy, 69978 Tel Aviv, Israel\Arefs{m}}
\item \Idef{triest_i}{Trieste Section of INFN, 34127 Trieste, Italy}
\item \Idef{triest}{University of Trieste, Department of Physics and Trieste Section of INFN, 34127 Trieste, Italy}
\item \Idef{triestictp}{Abdus Salam ICTP and Trieste Section of INFN, 34127 Trieste, Italy}
\item \Idef{turin_u}{University of Turin, Department of Physics and Torino Section of INFN, 10125 Turin, Italy}
\item \Idef{turin_i}{Torino Section of INFN, 10125 Turin, Italy}
\item \Idef{turin_p}{University of Eastern Piedmont, 15100 Alessandria,  and Torino Section of INFN, 10125 Turin, Italy}
\item \Idef{warsaw}{National Centre for Nuclear Research, 00-681 Warsaw, Poland\Arefs{n} }
\item \Idef{warsawu}{University of Warsaw, Faculty of Physics, 00-681 Warsaw, Poland\Arefs{n} }
\item \Idef{warsawtu}{Warsaw University of Technology, Institute of Radioelectronics, 00-665 Warsaw, Poland\Arefs{n} }
\item \Idef{yamagata}{Yamagata University, Yamagata, 992-8510 Japan\Arefs{k} }
\end{Authlist}
%
%
\vspace*{-\baselineskip}\renewcommand\theenumi{\alph{enumi}}
\begin{Authlist}
\item \Adef{a}{Also at IST, Universidade T\'ecnica de Lisboa, Lisbon, Portugal}
\item \Adef{bb}{Supported by the DFG Research Training Group Programme 1102  ``Physics at Hadron Accelerators''}
\item \Adef{b}{Also at Chubu University, Kasugai, Aichi, 487-8501 Japan\Arefs{k}}
\item \Adef{c}{Also at KEK, 1-1 Oho, Tsukuba, Ibaraki, 305-0801 Japan}
\item \Adef{cc}{Also at Moscow Institute of Physics and Technology, Moscow Region, 141700, Russia}
\item \Adef{y}{present address: National Science Foundation, 4201 Wilson Boulevard, Arlington, VA 22230, United States}
\item \Adef{x}{present address: RWTH Aachen University, III. Physikalisches Institut, 52056 Aachen, Germany}
\item \Adef{e}{Also at GSI mbH, Planckstr.\ 1, D-64291 Darmstadt, Germany}
\item \Adef{f}{Supported by the German Bundesministerium f\"ur Bildung und Forschung}
\item \Adef{g}{Supported by Czech Republic MEYS Grants ME492 and LA242}
\item \Adef{h}{Supported by SAIL (CSR), Govt.\ of India}
\item \Adef{i}{Supported by CERN-RFBR Grants 08-02-91009}
\item \Adef{j}{\raggedright Supported by the Portuguese FCT - Funda\c{c}\~{a}o para a Ci\^{e}ncia e Tecnologia, COMPETE and QREN, Grants CERN/FP/109323/2009, CERN/FP/116376/2010 and CERN/FP/123600/2011}
\item \Adef{k}{Supported by the MEXT and the JSPS under the Grants No.18002006, No.20540299 and No.18540281; Daiko Foundation and Yamada Foundation}
\item \Adef{l}{Supported by the DFG cluster of excellence `Origin and Structure of the Universe' (www.universe-cluster.de)}
\item \Adef{m}{Supported by the Israel Science Foundation, founded by the Israel Academy of Sciences and Humanities}
\item \Adef{n}{Supported by the Polish NCN Grant DEC-2011/01/M/ST2/02350}
\item [{\makebox[2mm][l]{\textsuperscript{*}}}] Deceased
\end{Authlist}

\clearpage
}

\maketitle

\section{Introduction}
Semi-Inclusive measurements of Deep Inelastic Scattering (SIDIS) of leptons off
nucleons provide information about the partonic structure of the nucleon and the
hadronization of partons, and hence offer a wide testing ground of Quantum
Chromodynamics (QCD). Subject of the present study are the transverse momentum
distributions of charged hadrons produced in the current fragmentation region in
lepton-nucleon scattering off unpolarised nucleons. The hadron transverse
momentum $p_T$ is defined with respect to the virtual photon direction. The
following standard notations are used: $\ell$ and $\ell '$ for the incoming and
outgoing lepton, $N$ for the target nucleon, $h$ for the outgoing hadron and $X$
for the unobserved particles in the final state; $l,l', P,$ and $p$ denote the
4-momenta of $\ell$, $\ell '$, $N$, and $h$. The general expression for the
differential SIDIS cross section describing the reaction $\ell+N \to \ell'+h+X$
in the one-photon approximation is~\cite{diehl,bacchetta}:
\begin{eqnarray}\label{SIDIS-az-dep}
  &&\frac{d^5\sigma^h(x_{Bj},Q^2,z,p_T^2,\phi_h)}{dx_{Bj} dQ^2dzdp_T^2d\phi_h} = 
\frac{d^4\sigma^h(x_{Bj},Q^2,z,p_T^2)}{2\pi dx_{Bj}dQ^2dzdp_T^2}\bigg(1+a_1(x_{Bj},Q^2,z,p_T^2)\cos\phi_h
  \nonumber \\
&&
\hspace{2cm} +a_2(x_{Bj},Q^2,z,p_T^2)\cos2\phi_h +\lambda a_3(x_{Bj},Q^2,z,p_T^2) \sin\phi_h\bigg)\,.
\end{eqnarray}
Here, $\lambda$ is the helicity of the incoming lepton and the standard SIDIS
variables are used: the 4-momentum transfer $q=(l-l')$, the photon virtuality
$Q^2=-q^2$, the Bjorken scaling variable $x_{Bj}=Q^2/2P\cdot q$, the hadron
fractional energy $z=P\cdot p/P\cdot q$ and the azimuthal angle $\phi_h$ of the
transverse momentum of the hadron with respect to the lepton scattering plane
around the virtual photon direction. After integration over $\phi_h$,
the cross section does not depend on the initial lepton polarisation
$\lambda$. The hadron multiplicity per interaction is defined as the ratio of
the differential SIDIS cross section over the differential DIS cross section $
d^2\sigma^{DIS}(x_{Bj},Q^2) / dx_{Bj} dQ^2 $. 
Thus, the differential hadron multiplicity, $d^2n^h /dz dp_T^2$, depends on four
variables, $x_{Bj},Q^2,z,p_T^2$:
\begin{equation}\label{mult-az-int}
  \frac{d^2n^h(x_{Bj},Q^2,z,p_T^2)}{dz dp_T^2} =
\frac{d^4\sigma^h(x_{Bj},Q^2,z,p_T^2)/(dx_{Bj} dQ^2 dz dp_T^2) }
{d^2\sigma^{DIS}(x_{Bj},Q^2)/(dx_{Bj} dQ^2)}\,.
\end{equation}

\begin{figure}
 \begin{center}
  \includegraphics[width=0.55\textwidth]{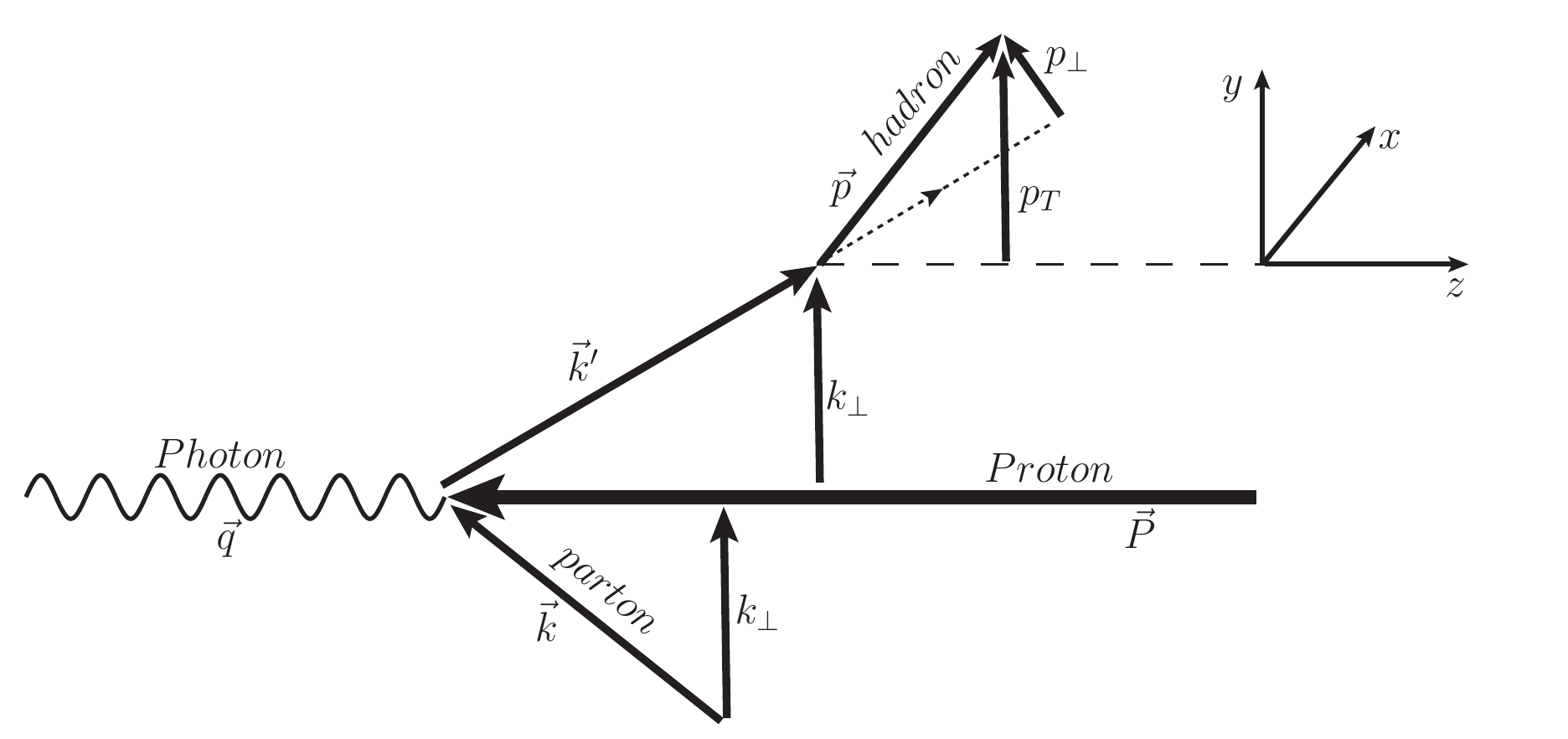}
 \end{center}
 \caption{Sketch showing the kinematic variables for the absorption of a
  virtual photon by a parton with intrinsic transverse momentum $k_{\perp}$
  and the subsequent hadronization. The transverse momentum of the observed
  hadron is denoted by $p_T$ when defined with respect to the virtual photon
  direction in the photon nucleon center of mass system and by $p_{\perp}$
  when defined with respect to the scattered parton direction.}
 \label{f_intrinsicKin}   
\end{figure}

Within a pQCD Leading Order (LO) parton model the shape of the $p_{T}^2$
distributions depends on the intrinsic transverse momentum $k_{\perp}$ of the
partons and the transverse momentum of the hadrons $p_{\perp}$ acquired during
parton fragmentation. The amount of the contributions of $k_{\perp}$ and
$p_{\perp}$ may depend on the hadron type, parton flavour, and on kinematic
variables such as $x_{Bj}$, $Q^2$ and $z$. Already in the 1970s, SIDIS was
understood as a tool to access the intrinsic transverse momentum of the partons
(see e.g.~\cite{Cahn:1978se} and references therein). The connection between the
intrinsic transverse momenta of the parton $k_{\perp}$ and that of the hadron
$p_{\perp}$ and the measured transverse momentum $p_{T}$ of the produced hadron
is illustrated in Fig.~\ref{f_intrinsicKin}, assuming single photon exchange and
leading order pQCD.
\label{s_intro}

During the last three decades significant efforts, both in experimental and
theoretical studies of (polarised) SIDIS, have been undertaken. Currently this
process is considered to be one of the most promising to study the hadronization
process and also the spin-dependent three-dimensional structure of the nucleon
(see, e.g.~\cite{Barone:2010zz}). Recently, a complete QCD treatment of
transverse momentum and spin-dependent SIDIS was presented
in Ref.~\cite{Collins:2011} where factorization was derived in terms of well defined
unintegrated or Transverse Momentum Dependent parton distribution and
fragmentation functions (TMDs) with individual hard scale evolution properties.
This formalism has been applied in Ref.~\cite{Aybat:2011zv} to
obtain the $Q^2$ evolution of unpolarised TMDs; a mandatory information needed
for a correct comparison of data measured in experiments at different hard
scales~\cite{Barone:2010zz}. 

Hadron leptoproduction has been studied by many experiments. Some recent
examples are: JLab~\cite{:2008rv}, HERMES~\cite{Airapetian:2001qk} and
E665~\cite{Adams:1997xb}. Earlier, EMC~\cite{Ashman:1991cj} covered most of the 
kinematic range of COMPASS. However, COMPASS has collected much more data in
this range and the statistical errors of the present analysis are therefore
significantly smaller, although only part of the available data has been used.
The results presented here are obtained from data taken during the year 2004.
More details of the analysis are described in Ref.~\cite{Rajotte:2010}.

\section{Experiment, Data Selection and Acceptance}
\begin{figure}
 \begin{center}
  \includegraphics[width=0.55\textwidth]{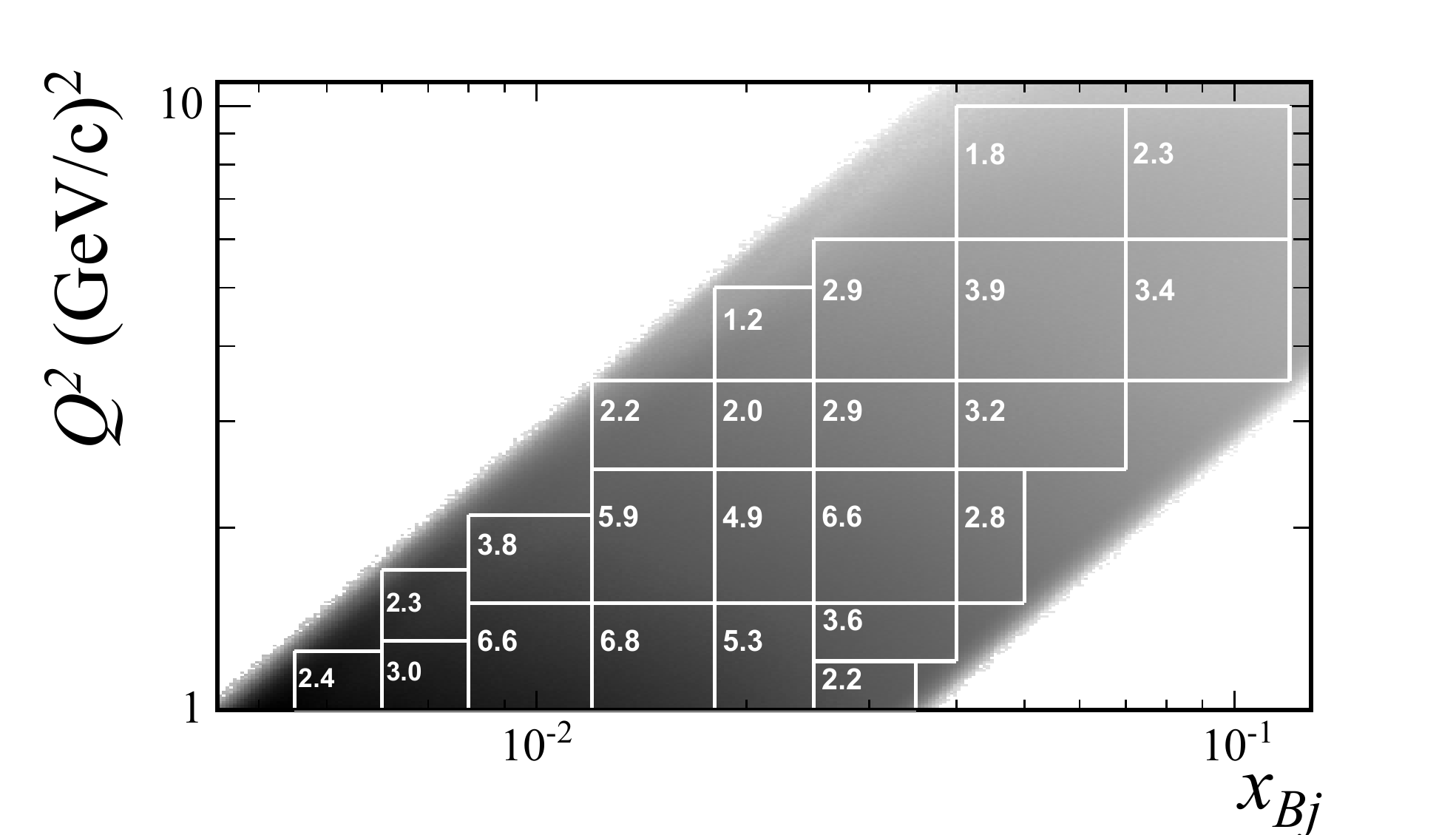}
 \end{center}
 \caption[Event distributions in the inclusive variables of the hadron
analysis] {Event distribution in the inclusive variables $Q^2$ and $x_{Bj}$ and
the 23 bins of the hadron cross section analysis. Within each bin, the fraction
of events contained is indicated in $\%$. }
\label{f_datadist_incl}
\end{figure}

The COMPASS experiment is installed on the M2 beam line of the CERN
SPS~\cite{Abbon:2007pq}. Polarised 160~\gom\ muons with an intensity of $2\times
10^8 \mu/$spill (one spill of $4.8$\,s length per $16.8$\,s) and a polarisation
of 80\% are scattered off a longitudinally polarised $^6$LiD target. In 2004 the
target consisted of two cells with opposite polarisation which was reversed
every 8 hours. It has been verified that summing up the data from both cells
yields a data sample with vanishing polarisation for the present analysis. The
COMPASS detector is a large acceptance two-stage spectrometer which covers the
kinematic range from quasi-real photoproduction to DIS. Both stages are equipped
with hadron calorimeters and use absorber walls for muon identification.
Charged particles emerging from the primary interaction vertex in the forward
direction are identified as muons if they traverse at least 30 radiation length,
otherwise they are identified as hadrons.  The selection requires reconstructed
trajectories in the detectors situated upstream and downstream of the first
magnet. This ensures that the track momentum and sign of charge are well defined
by bending in the magnetic field. The COMPASS ability to separate pions, kaons
and protons with a Ring Imaging Cherenkov detector was not used in this
analysis. Muon interactions with $Q^2 > 1.0$ \gomt\ and $0.1 < y < 0.9$ are
selected, where $y = \nu / E_{\mu}$, and $\nu = E_{\mu} -E_{\mu \prime} $ is the
difference between the laboratory energies of the incoming and outgoing muon
$\mu$ and $\mu ^{\prime}$. With the above selection, the hadronic energy squared
$W^2 = 2M\nu + M^2 - Q^2$ is $>25$ \gom, above the nucleon resonance
region. Here, $M$ is the nucleon mass. The 
total number of inclusive events selected for this analysis is $ 45.8\times
10^6$, corresponding to an integrated luminosity of 775 pb$^{-1}$.  The events
are sampled into 23 intervals in $Q^2$ from 1 to 10 \gomt\ and $x_{Bj} $ from
0.004 to 0.12, as shown in Fig.~\ref{f_datadist_incl}. The ranges and average
values of $Q^2$ and $x_{Bj} $ are shown in Tab.\,\ref{Table1}. Each of these
$(x_{Bj}, Q^2)$ intervals is further subdivided into 8 intervals in $z$ from 0.2
to 0.8.

In order to correct for event losses caused by the non uniform
acceptance of the COMPASS spectrometer, a full Monte Carlo (MC) simulation has
been performed. The events were generated with LEPTO~\cite{LEPTO}, passed
through the spectrometer with a GEANT~\cite{GEANT3} based simulation program
and reconstructed with the reconstruction software as the real data events.

\begin{figure}
 \begin{center}
  \includegraphics[width=0.55\textwidth]{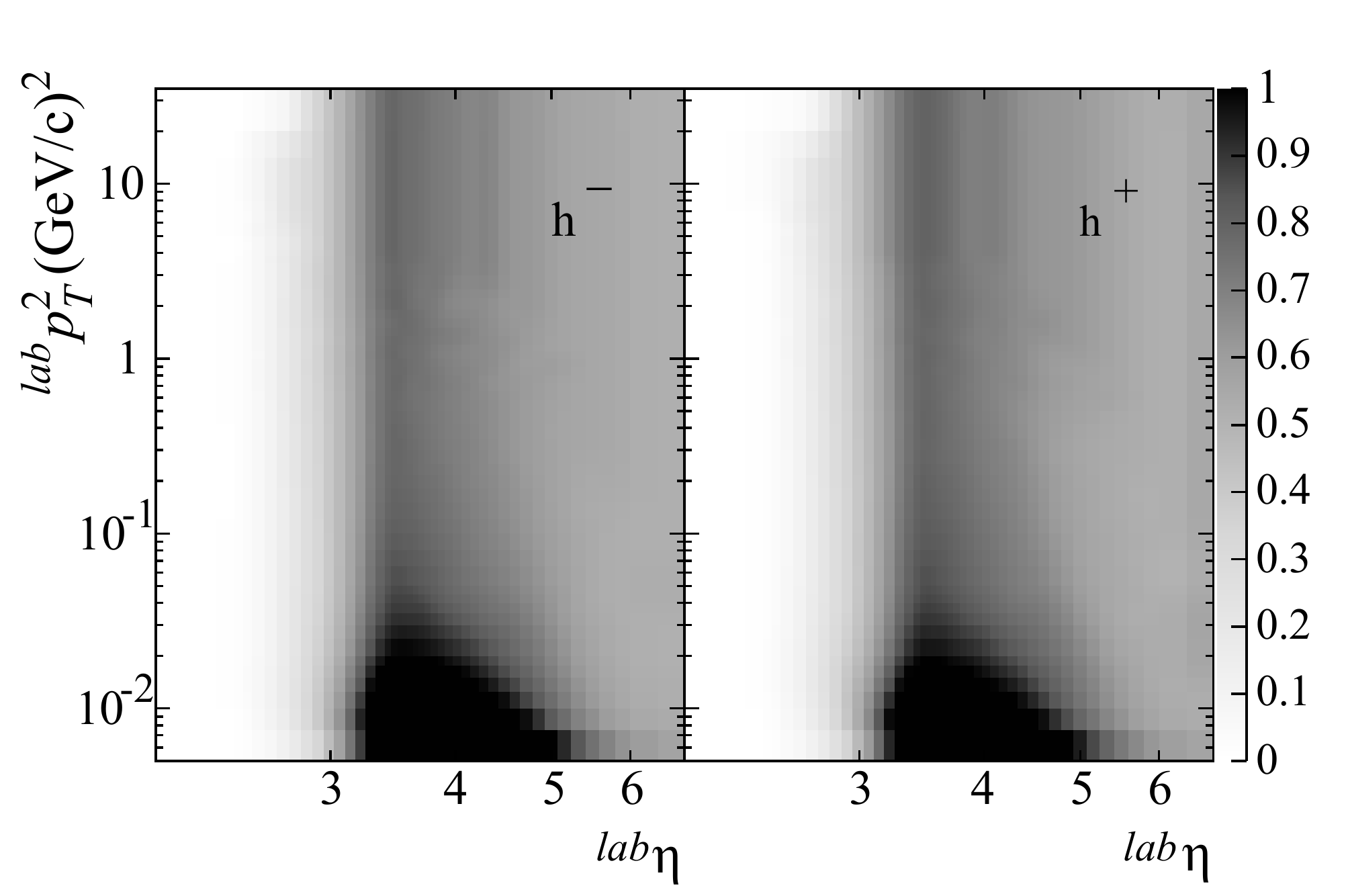}
 \end{center}
 \caption{Hadron acceptances $A_{h^-}$ and $A_{h^+}$ determined with the Monte
  Carlo simulation for $Q^2>1$ \gomt\ as a function of $^{lab}p_T$ and
  $^{lab}\eta$ for negative hadrons $h^-$ (left) and positive hadrons $h^+$
  (right). The acceptances have been smoothed in order to reduce the
  granularity from the binning. }
 \label{f_acceptance_tables}
\end{figure}

The SIDIS acceptances $A^{(+,-)}_{SIDIS}$ for detecting, together with the
scattered muon, a positive ($h^+$) or negative hadrons ($h^-$) respectively
factorize in an inclusive muon acceptance $A_{incl}(Q^2,y)$ and a positive or
negative hadron acceptance $A_{h^{(+,-)}}(^{lab}p_{T},\leftexp{lab}{\eta})$.
These acceptances depend on the spectrometer characteristics, making the use of
variables defined in the laboratory frame preferable; therefore, the transverse momentum
$^{lab} p_T$, the polar angle $^{lab} \theta$, and the pseudorapidity
$^{lab}\eta = -\ln (tan\frac{^{lab}\theta}{2})$ of the hadron are defined with
respect to the direction of the incoming muon. The choice of $^{lab} \theta$ is
particularly convenient to exhibit the acceptance cut due to the aperture limit
of the polarised target magnet at $^{lab} \theta = 70 $\,mrad for the upstream
edge of the target.  The factorization of hadron and muon acceptances implies
that the differential multiplicities only depend on $A_{h^{(+,-)}}$ since
$A_{incl}$ cancels, see Eq.~\ref{mult-az-int}. Figure~\ref{f_acceptance_tables} shows
the hadron acceptances $A_{h^-}$ and $A_{h^+}$ used in the analysis.

The four-dimensional acceptance used in the present analysis is integrated over
the azimuthal angle of the hadrons, i.e. does not take into account the
azimuthal modulations in the cross section~\cite{bacchetta}. The systematic
effect on the extracted $\langle p_T^2 \rangle$  have been investigated and
found to be negligible. 

\begin{table}[htdp]
\centering
\begin{tabular}{|p{1.cm}|p{1.5cm}p{1.5cm}p{1.5cm}|p{1.5cm}p{1.5cm}p{1.5cm}|} \hline
 Bin & $x_{bj}^{min}$ & $x_{bj}^{max}$ & $\langle x_{bj} \rangle$ & $Q^2_{min}$ & $Q^2_{max}$ & $\langle Q^2 \rangle$ \\ 
 \hline
 1  & 0.0045 & 0.0060 & 0.0052 & 1.0 & 1.25  & 1.11 \\ 
 2  & 0.0060 & 0.0080 & 0.0070 & 1.0 & 1.30  & 1.14 \\ 
 3  & 0.0060 & 0.0080 & 0.0070 & 1.3 & 1.70  & 1.48 \\ 
 4  & 0.0080 & 0.0120 & 0.0099 & 1.0 & 1.50  & 1.22 \\ 
 5  & 0.0080 & 0.0120 & 0.0099 & 1.5 & 2.10  & 1.76 \\ 
 6  & 0.0120 & 0.0180 & 0.0148 & 1.0 & 1.50  & 1.22 \\ 
 7  & 0.0120 & 0.0180 & 0.0148 & 1.5 & 2.50  & 1.92 \\ 
 8  & 0.0120 & 0.0180 & 0.0150 & 2.5 & 3.50  & 2.90 \\ 
 9  & 0.0180 & 0.0250 & 0.0213 & 1.0 & 1.50  & 1.23 \\ 
 10  & 0.0180 & 0.0250 & 0.0213 & 1.5 & 2.50  & 1.92 \\ 
 11  & 0.0180 & 0.0250 & 0.0213 & 2.5 & 3.50  & 2.94 \\ 
 12  & 0.0180 & 0.0250 & 0.0216 & 3.5 & 5.00  & 4.07 \\ 
 13  & 0.0250 & 0.0350 & 0.0295 & 1.0 & 1.20  & 1.10 \\ 
 14  & 0.0250 & 0.0400 & 0.0316 & 1.2 & 1.50  & 1.34 \\ 
 15  & 0.0250 & 0.0400 & 0.0318 & 1.5 & 2.50  & 1.92 \\ 
 16  & 0.0250 & 0.0400 & 0.0319 & 2.5 & 3.50  & 2.95 \\ 
 17  & 0.0250 & 0.0400 & 0.0323 & 3.5 & 6.00  & 4.47 \\ 
 18  & 0.0400 & 0.0500 & 0.0447 & 1.5 & 2.50  & 1.93 \\ 
 19  & 0.0400 & 0.0700 & 0.0533 & 2.5 & 3.50  & 2.95 \\ 
 20  & 0.0400 & 0.0700 & 0.0536 & 3.5 & 6.00  & 4.57 \\ 
 21  & 0.0400 & 0.0700 & 0.0550 & 6.0 & 10.0  & 7.36 \\ 
 22  & 0.0700 & 0.1200 & 0.0921 & 3.5 & 6.00  & 4.62 \\ 
 23  & 0.0700 & 0.1200 & 0.0932 & 6.0 & 10.0  & 7.57 \\
\hline
\end{tabular}
\caption{Definition of the 23 bins of $x_{bj}$ and $Q^{2}$ and corresponding
  mean values; $Q^2$ is in units of \gomt.
}
\label{Table1}
\end{table}

\section{Results}
The differential multiplicities $d^2 n^{h \pm}/dzd p_{T}^2 $ in bins of $(Q^2,\
x_{Bj})$ are defined in the introduction in terms of the 
semi-inclusive and inclusive differential cross sections. They are obtained as
the acceptance corrected number of hadrons $ \Delta^4 N^{h \pm}$ in $8 \times
40$ ($z$, $p_{T}^2$) bins and 23 ($\Delta x_{Bj}, \Delta Q^2$) bins, divided
by the number $ \Delta^2 N^{\mu}$ of muon interactions in the same ($\Delta
x_{Bj},\Delta Q^2$) bins:
\begin{align}
   \frac{d^2 n^{h \pm}(z,p_T^2,x_{Bj},Q^2)}{dzd p_{T}^2}\bigg|_{\Delta x_{Bj}
 \Delta Q^2} \approx \frac{\Delta^4 N^{h \pm}(z,p_T^2,x_{Bj},Q^2)/(\Delta z
 \Delta p_{T}^2 \Delta x_{Bj} \Delta Q^2)}{\Delta^2 N^{\mu}(x_{Bj}, Q^2) /
 (\Delta x_{Bj} \Delta Q^2)}.
\label{eq_xs_mult}
\end{align}

\begin{figure}
 \begin{center}
  \includegraphics[width=1.0\textwidth]{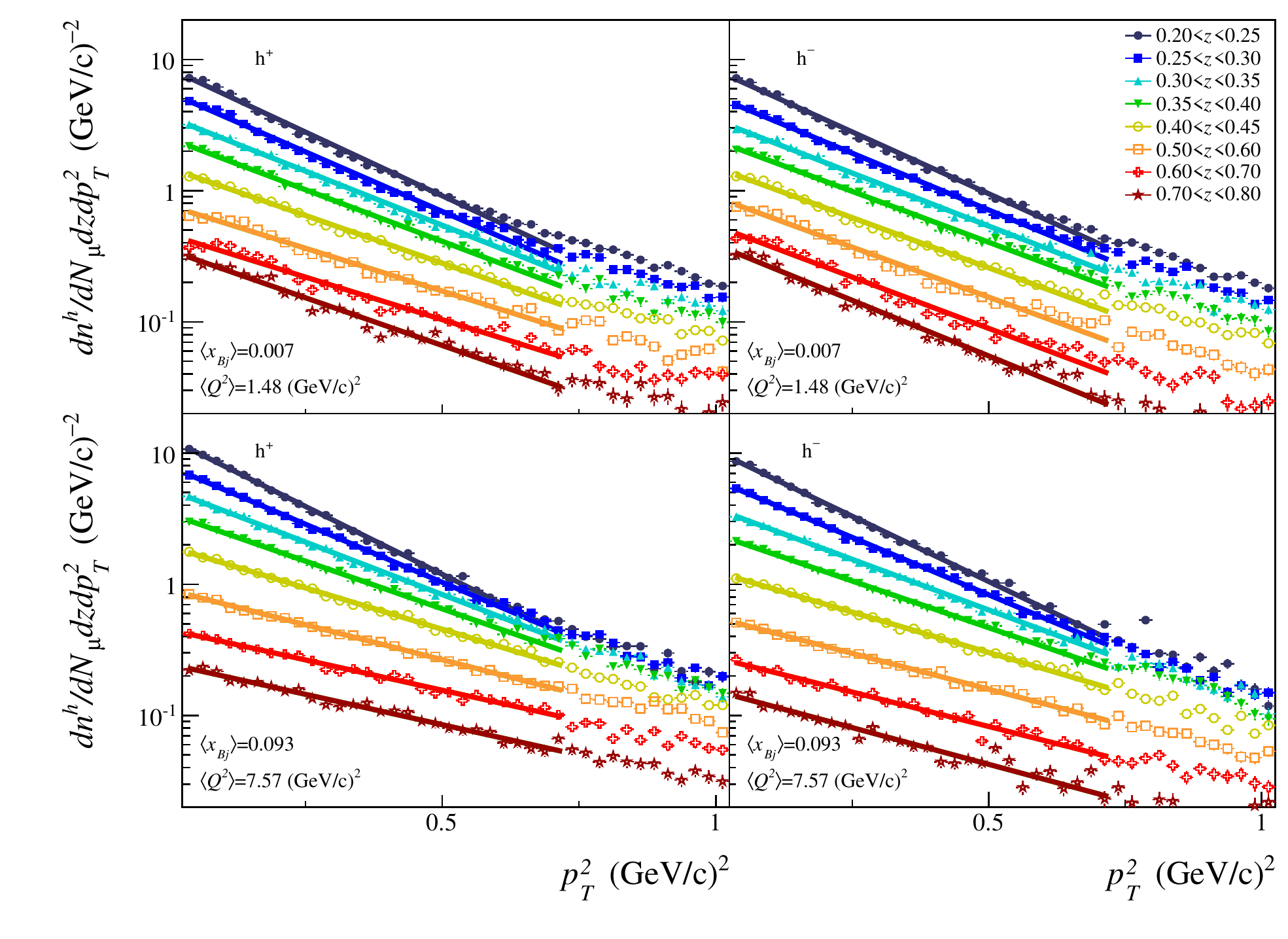}
 \end{center}
 \caption{The $p_T^2$ dependence of the differential multiplicities $d^2 n^{h
}/dzd p_{T}^2 $ of positive hadrons (left) and negative hadrons (right) fitted
by an exponential for $1<Q^2\ \gomt <1.5$, $0.006<x_{Bj}<0.008$ (top) and
$6<Q^2\ \gomt <10$, $0.07<x_{Bj}<0.12$ (bottom) subdivided into eight $z$
intervals, see legend of upper pictures. The average values $\langle Q^2
\rangle$ and $\langle x_{Bj} \rangle$ for the chosen ($Q^2, x_{Bj}$) intervals
are indicated in the pictures. The systematic error of 5\% is not included in
the errors.}
\label{f_multNfit}
\end{figure}

The distributions for two selected ($Q^2$, $x_{Bj}$) bins are shown in
Fig.~\ref{f_multNfit} for all $z$ intervals. The full data set, including
more $p_T^2$ bins, is available on HEPDATA~\cite{HEPDATA}. As can be seen
from Eq.~\ref{eq_xs_mult} the uncertainty of the integrated luminosity cancels and the
only contributions to systematic uncertainties of the multiplicities come from
the hadron acceptance and the assumption of factorization of hadron and muon
acceptance. The total systematic uncertainty due to acceptance has been
estimated to be 5\%~\cite{Rajotte:2010}. Only statistical errors are shown
in the figures.

\begin{figure}
 \begin{center}
  \includegraphics[width=1.0\textwidth]{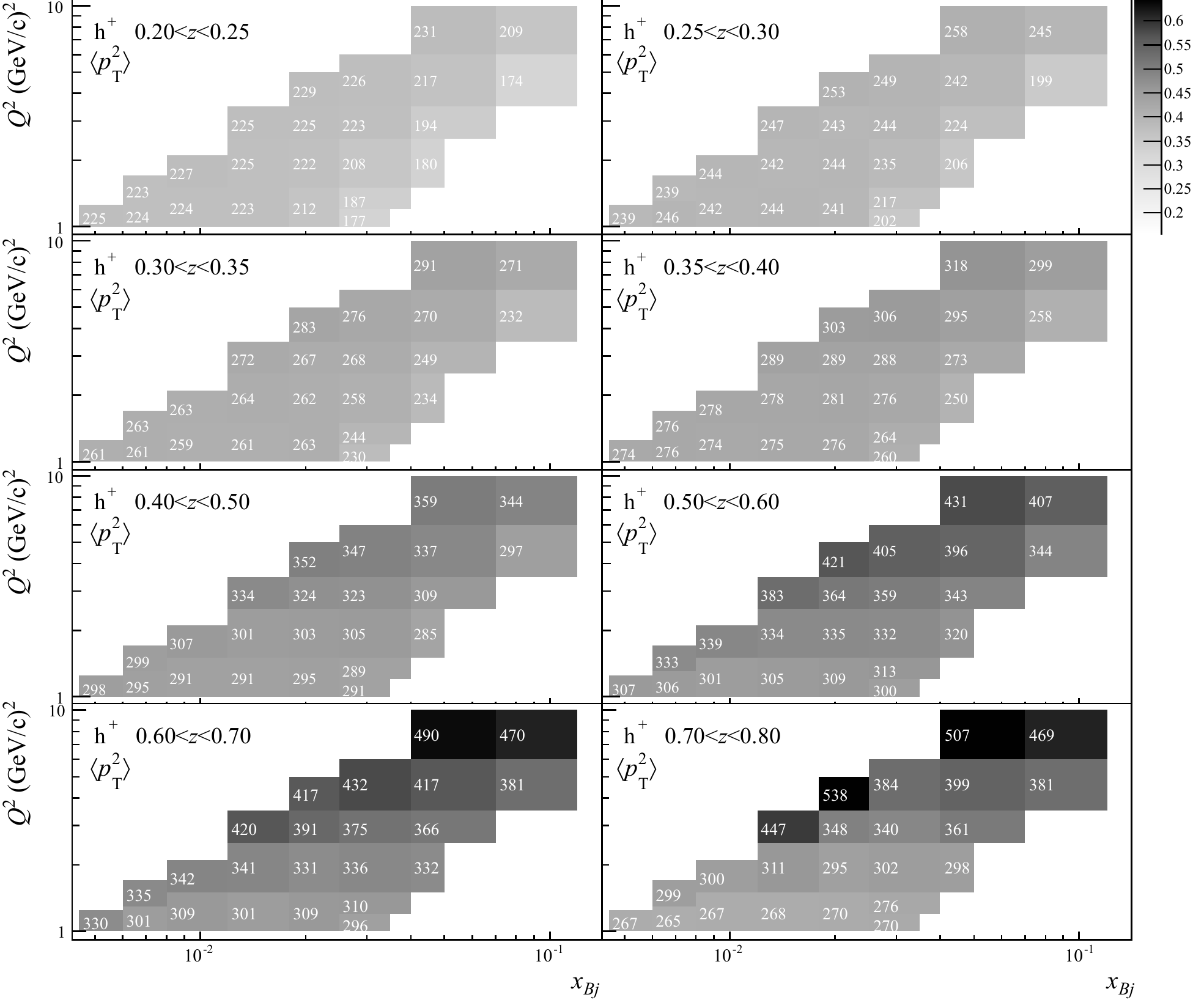}
\end{center}
\caption{Fitted $\langle p_T^2 \rangle$ vs ($x_{Bj}$, $Q^2$) for all $z$
intervals for positive hadrons. The values are both written inside each
interval and shown as a gray scale. The same gray scale is used for all the  
plots. The written values are in units of $\gomt \times 1000$.}
\label{f_pt2VSxQp}
\end{figure}

\begin{figure}
 \begin{center}
  \includegraphics[width=1.0\textwidth]{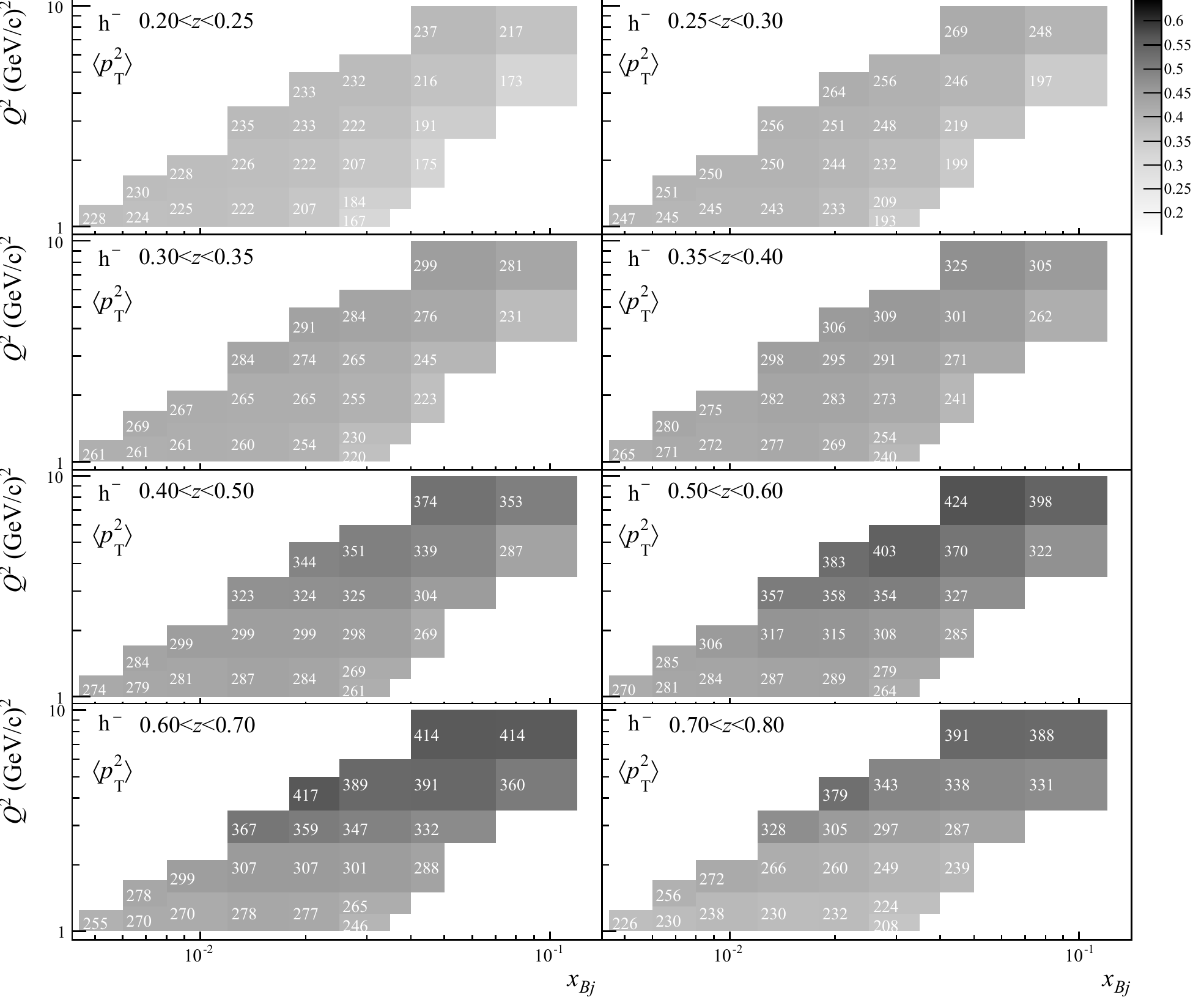}
\end{center}
\caption{ As Fig.~\protect\ref{f_pt2VSxQp} but for negative hadrons. }
\label{f_pt2VSxQn}
\end{figure}

The fits are performed at values of $p_T$ smaller than 0.85 \gom\ to
stay away from pQCD effects where the assumption of a simple exponential
distribution is known to fail~\cite{Anselmino:2005nn, Anselmino:2006rv} and at
$p_T$ larger than 0.1 \gom\ to exclude a region where the experimental
resolution may affect the distribution. In this range, the $ p_T^2$
distributions are fitted with a single exponential functions $ A
e^{-p_T^2/\langle p_T^2 \rangle}$ to extract the inverse slope $\langle p_T^2
\rangle$. The values of
$\langle p_T^2 \rangle$ for all intervals of $x_{Bj}$, $Q^2$ and $z$ are shown
in Figs.~\ref{f_pt2VSxQp} and \ref{f_pt2VSxQn} and in Tabs.~\ref{Table2} and \ref{Table3}.  These
figures and tables contain the basic experimental information extracted from the
fits of the $p_T^2 $ distributions.


\begin{sidewaystable}[htdp]
\centering
\begin{tabular}{|p{1.2cm}|p{1.7cm}p{1.7cm}p{1.7cm}p{1.7cm}p{1.7cm}p{1.7cm}p{1.7cm}p{1.7cm}|} \hline
$Q^2,x_{Bj}$ & $ z $ bins &$\rightarrow$  &  &  &   &    &   &    \\
Bin & 0.2$\div$0.25 &  0.25$\div$0.3 & 0.3$\div$0.35 & 0.35$\div$0.4 &  0.4$\div$0.5  & 0.5$\div$0.6    &  0.6$\div$0.7   & 0.7$\div$0.8   \\
\hline
1 & .2254(13)  &  .2394(17) & .2607(23) & .2738(30)  &  .2981(31)  &  .3072(43)  &  .3299(65)  & .2674(57)   \\
2 & .2241(10)  &  .2457(14) & .2614(19) & .2763(24)  &  .2952(24)  &  .3060(35)  &  .3009(44)  & .2650(46)   \\
3 & .2234(13)  &  .2390(18) & .2631(24) & .2761(31)  &  .2986(32)  &  .3327(52)  &  .3348(69)  & .2985(68)   \\
4 & .2239( 6)  &  .2424( 9) & .2592(12) & .2738(15)  &  .2915(15)  &  .3006(21)  &  .3087(29)  & .2670(30)   \\
5 & .2270( 9)  &  .2438(13) & .2634(17) & .2777(22)  &  .3073(24)  &  .3387(38)  &  .3419(51)  & .2998(55)   \\
6 & .2231( 6)  &  .2442( 8) & .2610(11) & .2755(14)  &  .2914(13)  &  .3053(19)  &  .3010(25)  & .2677(27)   \\
7 & .2246( 7)  &  .2418( 9) & .2641(13) & .2779(17)  &  .3013(17)  &  .3338(27)  &  .3410(38)  & .3105(44)   \\
8 & .2250(13)  &  .2470(18) & .2723(25) & .2893(33)  &  .3338(38)  &  .3829(67)  &  .420(11 )  & .447(17 )  \\
9 & .2122( 7)  &  .2408( 9) & .2630(12) & .2762(15)  &  .2953(15)  &  .3087(21)  &  .3094(26)  & .2699(27)   \\
10 & .2222( 7)  &  .2439(10) & .2617(13) & .2806(17)  &  .3034(17)  &  .3347(28)  &  .3309(37)  & .2946(39)   \\
11 & .2250(12)  &  .2425(16) & .2667(22) & .2886(30) &  .3243(34) &  .3642(57)  &  .3912(88)  & .3485(95)   \\
12 & .2289(17)  &  .2530(24) & .2828(35) & .3031(48) &  .3515(55) &  .421(11 )  &  .417(14 )  & .538(34 )  \\
13 & .1766(13)  &  .2020(15) & .2299(18) & .2597(23) &  .2909(22) &  .3003(29)  &  .2956(34)  & .2704(37)  \\
14 & .1866( 9)  &  .2171(13) & .2436(15) & .2640(18) &  .2893(17) &  .3131(24)  &  .3099(30)  & .2758(31)  \\
15 & .2078( 6)  &  .2355( 8) & .2577(11) & .2759(14) &  .3050(14) &  .3319(22)  &  .3364(29)  & .3025(31)  \\
16 & .2229( 9)  &  .2441(13) & .2678(18) & .2882(23) &  .3230(26) &  .3587(42)  &  .3749(59)  & .3395(69)  \\
17 & .2257(10)  &  .2493(14) & .2761(20) & .3064(28) &  .3468(32) &  .4050(59)  &  .4321(91)  & .384(10 )  \\
18 & .1799(11)  &  .2063(13) & .2340(17) & .2500(19) &  .2853(20) &  .3197(30)  &  .3321(40)  & .2984(42)  \\
19 & .1944( 9)  &  .2245(12) & .2486(15) & .2735(20) &  .3088(22) &  .3434(34)  &  .3656(49)  & .3609(62)  \\
20 & .2167( 8)  &  .2415(11) & .2700(15) & .2947(21) &  .3370(23) &  .3959(42)  &  .4170(63)  & .3994(76)  \\
21 & .2311(13)  &  .2579(18) & .2908(27) & .3178(37) &  .3588(41) &  .4307(80)  &  .490(14 )  & .507(21 ) \\
22 & .1738(10)  &  .1990(12) & .2319(16) & .2578(21) &  .2969(22) &  .3437(38)  &  .3809(57)  & .3809(74)  \\
23 & .2091(10)  &  .2448(15) & .2714(21) & .2989(28) &  .3441(31) &  .4072(57)  &  .470(10 )  & .469(13 ) \\
\hline
\end{tabular}
\caption{Fitted $\langle p_T^2 \rangle$ in units of \gomt\ for the 23
($x_{Bj}$, $Q^2$) bins (rows) and the 8 $z$ intervals (columns) for positive
hadrons. The error of the least significant digit(s) is given in parentheses.
Same information as Fig.~\protect\ref{f_pt2VSxQp}}
\label{Table2}
\end{sidewaystable}

\begin{sidewaystable}[htdp]
\centering
\begin{tabular}{|p{0.5cm}|p{1.7cm}p{1.7cm}p{1.7cm}p{1.7cm}p{1.7cm}p{1.7cm}p{1.7cm}p{1.7cm}|} \hline
Bin & 0.2$\div$0.25 &  0.25$\div$0.3 & 0.3$\div$0.35 & 0.35$\div$0.4 &  0.4$\div$0.5  & 0.5$\div$0.6    &  0.6$\div$0.7   & 0.7$\div$0.8   \\
\hline
1  &  .2285(13)  &  .2472(18)  &  .2606(24)  &  .2651(29)  &  .2741(27)  &  .2699(35)  & .2551(41)  &  .2259(42)  \\
2  &  .2241(10)  &  .2450(15)  &  .2607(20)  &  .2713(25)  &  .2793(23)  &  .2806(31)  & .2697(37)  &  .2297(37)  \\
3  &  .2305(14)  &  .2510(19)  &  .2690(25)  &  .2802(33)  &  .2843(30)  &  .2851(40)  & .2778(50)  &  .2562(54)  \\
4  &  .2254( 7)  &  .2448( 9)  &  .2614(13)  &  .2719(16)  &  .2814(15)  &  .2836(20)  & .2700(24)  &  .2375(26)  \\
5  &  .2276(10)  &  .2498(14)  &  .2666(19)  &  .2752(23)  &  .2987(24)  &  .3056(33)  & .2990(42)  &  .2722(46)  \\
6  &  .2216( 7)  &  .2431( 9)  &  .2604(12)  &  .2766(15)  &  .2869(14)  &  .2870(19)  & .2780(24)  &  .2297(23)  \\
7  &  .2260( 7)  &  .2498(10)  &  .2653(14)  &  .2823(18)  &  .2989(18)  &  .3171(27)  & .3068(35)  &  .2657(37)  \\
8  &  .2346(14)  &  .2555(19)  &  .2841(28)  &  .2976(36)  &  .3233(38)  &  .3572(62)  & .3674(88)  &  .328(10 )  \\
9  &  .2071( 8)  &  .2326(10)  &  .2540(13)  &  .2686(16)  &  .2837(15)  &  .2888(20)  & .2767(24)  &  .2316(24)  \\
10 &  .2219( 8)  &  .2440(11)  &  .2646(14)  &  .2831(19)  &  .2992(19)  &  .3150(28)  & .3068(36)  &  .2596(35)  \\
11 &  .2331(13) &  .2512(18) &  .2742(25) &  .2953(34) &  .3243(37) &  .3578(62) &  .3587(85)  &  .3048(87)  \\
12 &  .2333(18) &  .2644(27) &  .2908(40) &  .3057(50) &  .3443(58) &  .383(10 ) &  .417(16 ) &   .379(18 )  \\
13 &  .1669(12) &  .1933(16) &  .2201(20) &  .2400(23) &  .2614(21) &  .2640(27) &  .2457(29)  &  .2083(28)  \\
14 &  .1835(11) &  .2086(12) &  .2302(15) &  .2536(19) &  .2693(18) &  .2794(24) &  .2648(27)  &  .2240(26)  \\
15 &  .2067( 7) &  .2316( 9) &  .2548(12) &  .2728(15) &  .2976(16) &  .3081(22) &  .3015(27)  &  .2492(26)  \\
16 &  .2219(10) &  .2482(14) &  .2652(19) &  .2907(26) &  .3254(29) &  .3541(47) &  .3473(62)  &  .2971(60)  \\
17 &  .2325(11) &  .2556(16) &  .2844(23) &  .3092(31) &  .3506(37) &  .4033(68) &  .3890(87)  &  .3432(95)  \\
18 &  .1746(11) &  .1994(14) &  .2232(18) &  .2411(21) &  .2692(21) &  .2850(30) &  .2882(38)  &  .2394(35)  \\
19 &  .1915( 9) &  .2193(13) &  .2452(18) &  .2715(23) &  .3040(25) &  .3268(39) &  .3319(51)  &  .2869(52)  \\
20 &  .2163( 9) &  .2456(12) &  .2765(18) &  .3007(25) &  .3389(28) &  .3703(45) &  .3909(70)  &  .3382(73)  \\
21 &  .2370(14) &  .2686(21) &  .2994(31) &  .3246(44) &  .3744(53) &  .4235(96) &  .414(13 ) &   .391(17 ) \\
22 &  .1730(10) &  .1967(13) &  .2314(19) &  .2623(27) &  .2867(26) &  .3218(41) &  .3598(66)  &  .3310(75)  \\
23 &  .2166(13) &  .2482(17) &  .2814(25) &  .3050(35) &  .3531(41) &  .3976(71) &  .414(11 ) &   .388(13 ) \\
\hline
\end{tabular}
\caption{ Fitted $\langle p_T^2 \rangle$ in units of \gomt\ for the 23
($x_{Bj}$, $Q^2$) bins (rows) and the 8 $z$ intervals (columns) for negative
hadrons. The error of the least significant digit(s) is given in parentheses.
Same information as Fig.~\protect\ref{f_pt2VSxQn}}
\label{Table3}
\end{sidewaystable}

\begin{figure}
 \begin{center}
  \includegraphics[width=0.90\textwidth]{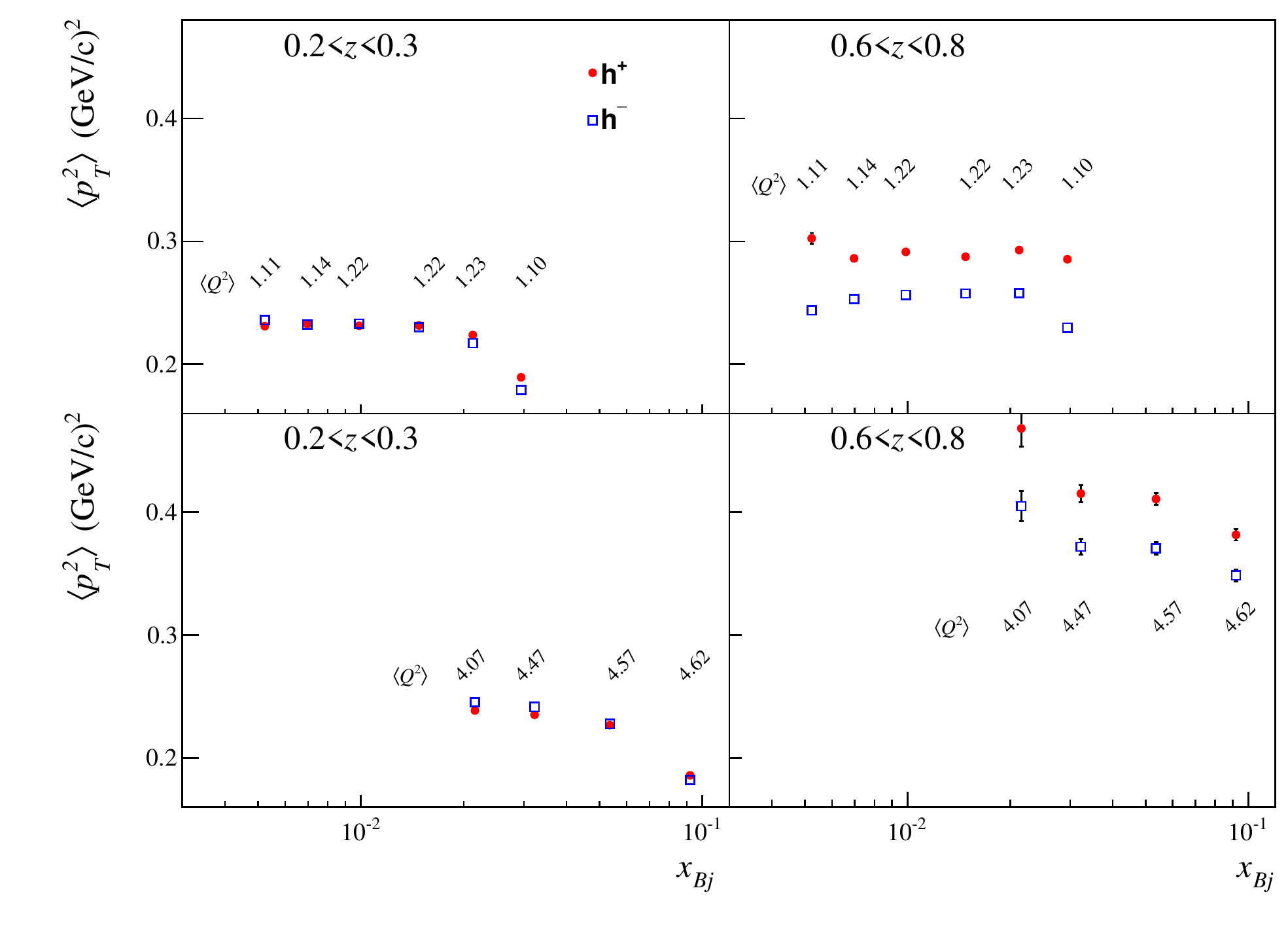}
 \end{center}
\caption{The fitted $\langle p_T^2 \rangle$ vs $x_{Bj}$ for two different $Q^2$
intervals (top and bottom) and for a low-$z$ bin ($0.2 < z < 0.3$), left, and a
high-$z$ bin ($0.6 < z < 0.8$), right, for positive and negative hadrons (red
filled circles and blue open boxes). In these figures, the $\langle p_T^2
\rangle$ values are obtained from a fit over $0.1 < p_T<0.85$ \gom. The
average value $\langle Q^2 \rangle$ for each $x_{Bj}$ bin is indicated.}
\label{f_pt2VSx}
\end{figure}

In Fig.~\ref{f_pt2VSx} the dependence of the fitted $\langle p_T^2 \rangle$ on
$x_{Bj}$ is shown for a low-$z$ and a high-$z$ bin and for a low- and a
high-$Q^2$ bin. At higher $z$ the positive hadrons clearly have higher
$\langle p_T^2 \rangle$ than the negative hadrons. For hadrons with lower $z$
however, no difference is observed in the $p_T^2$ distributions. A similar
behaviour was already reported by HERMES~\cite{Jgoun:2001ck} for the average
$p_T^2$, not determined by a fit but from a standard average over the entire
$p_T$ range, i.e. $\langle p_T^2 \rangle_{all}$. The $z$-dependence as well as
the hadron charge dependence of the $p_T^2$ distributions will be further
investigated below and is related to the intrinsic transverse momentum of the
partons.

\begin{figure}
 \begin{center}
  \includegraphics[width=0.90\textwidth]{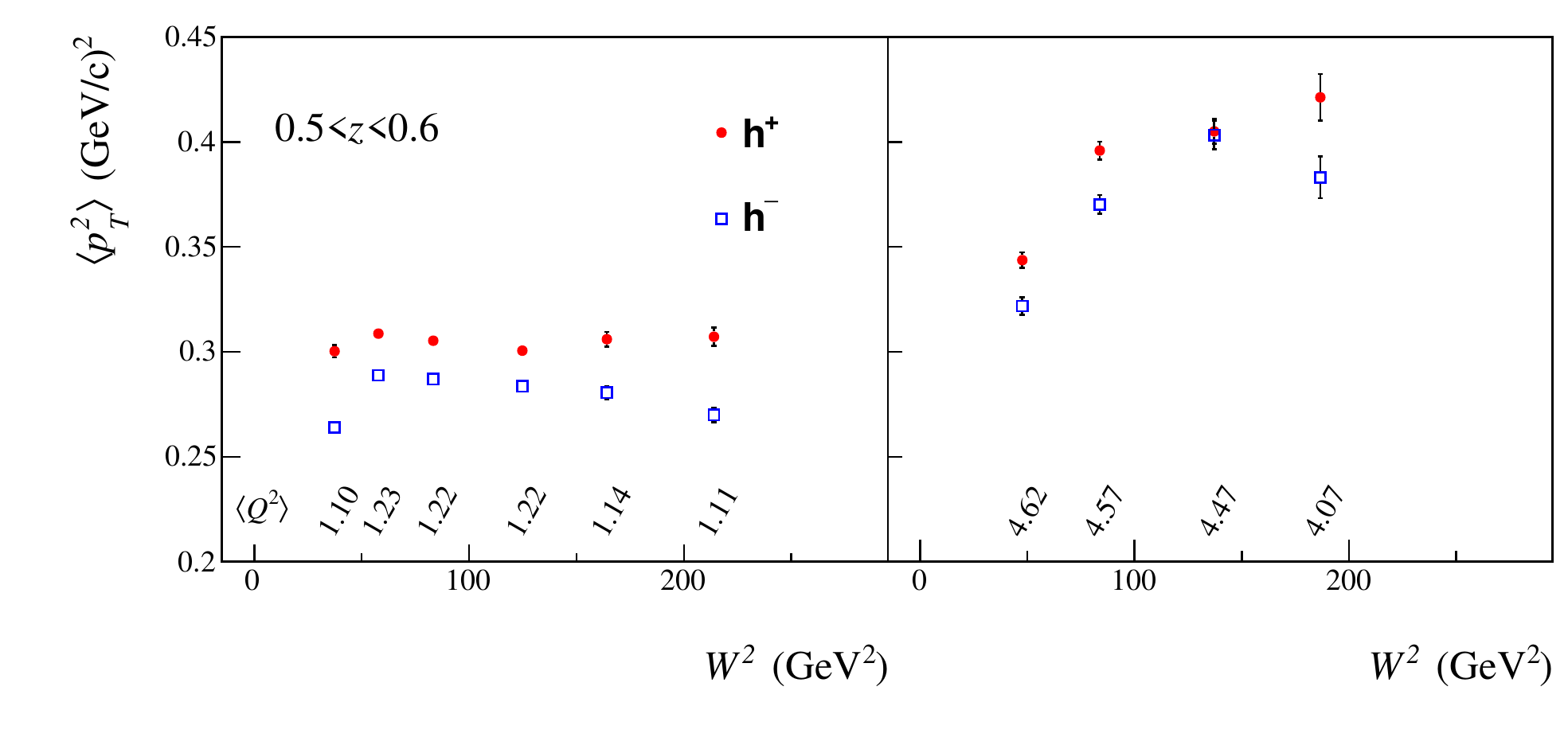}
 \end{center}
\caption{The fitted $\langle p_T^2 \rangle$ vs $W^2$ for $0.5 <z <0.6$ and for a
low (left) and a high (right) $Q^2$ interval, from a fit over $0.1 <
p_T\ \gom<0.85$. This is to be compared with Fig.~\protect\ref{f_avptVSw2} where
$\langle p_T^2 \rangle_{all}$ is plotted. The average $\langle Q^2 \rangle$ for
each $W^2$ bin are indicated.}
\label{f_pt2VSw2}
\end{figure}

\begin{figure}
 \begin{center}
  \includegraphics[width=0.90\textwidth]{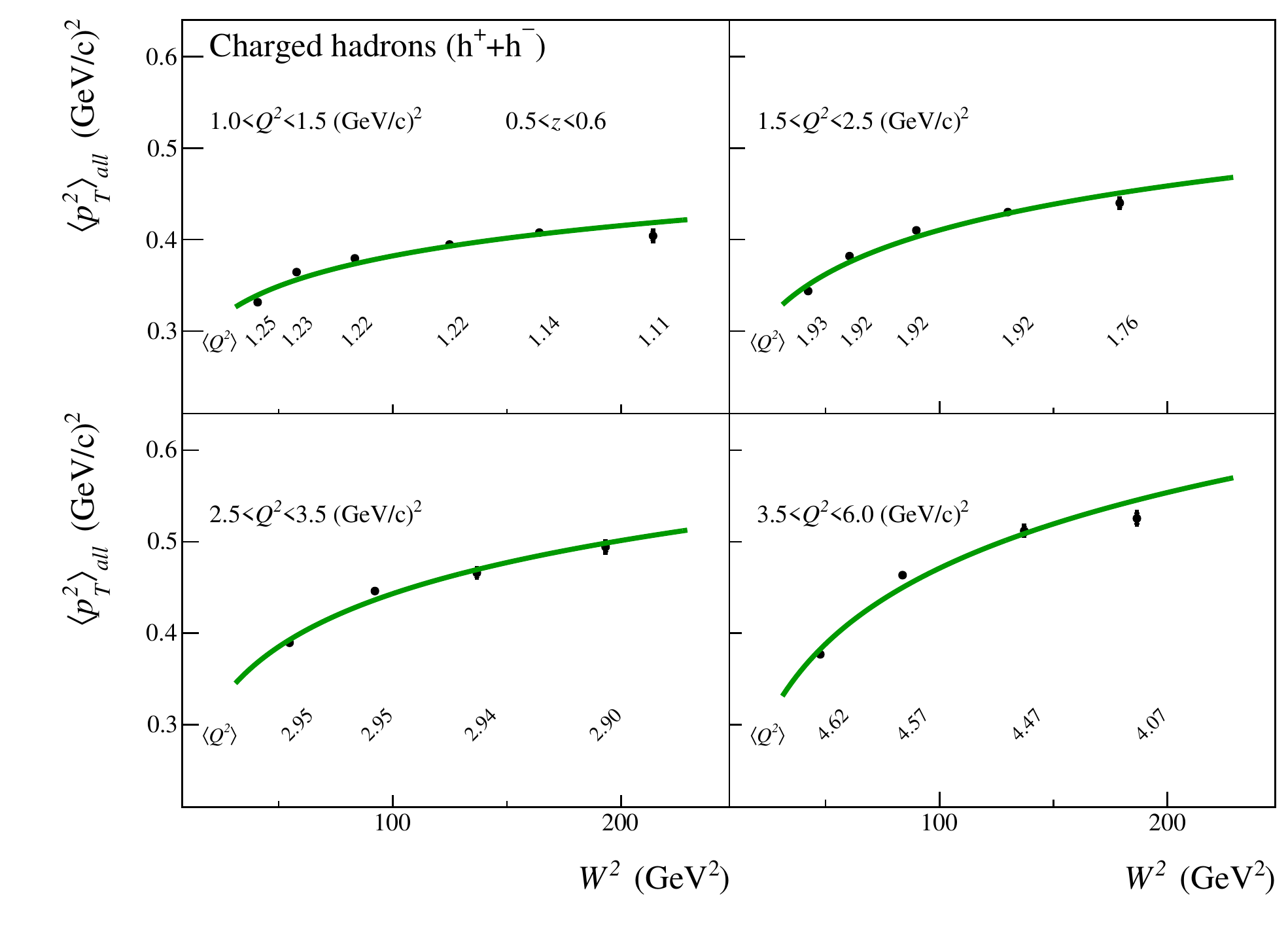} 
 \end{center}
\caption{Statistical average $\langle p_T^2 \rangle_{all}$
over the entire $p_T$ range for charged hadrons ($h^+$ and $h^-$ summed up) as a function of $W^2$,
for $0.5<z<0.6$ and four $Q^2$ intervals, indicated in the figures.
The green lines represent fits where a linear function of $\ln W^2$ was assumed.}
\label{f_avptVSw2}
\end{figure}

\begin{figure}
 \begin{center}
  \includegraphics[width=0.48\textwidth]{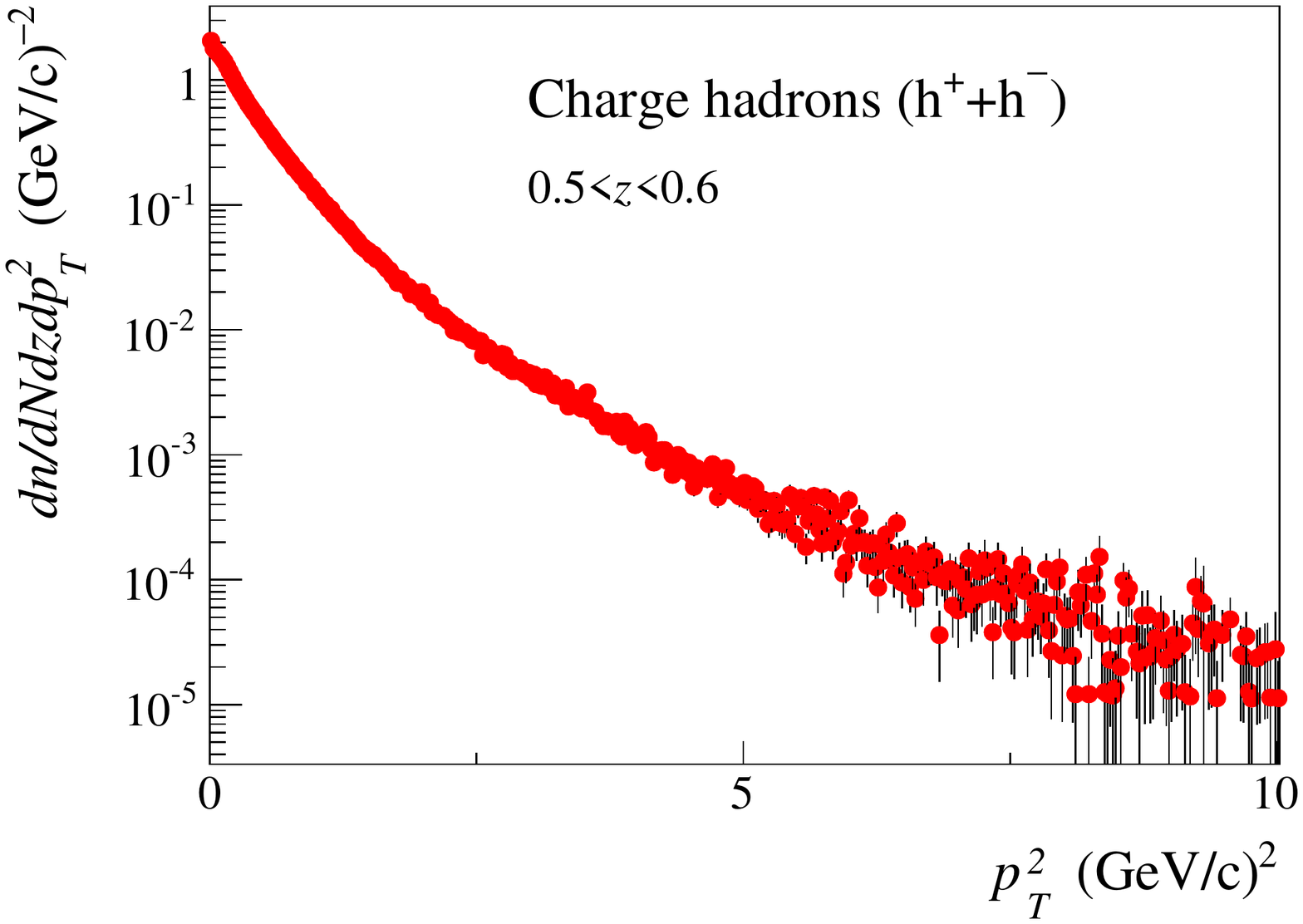} \hfill
  \includegraphics[width=0.48\textwidth]{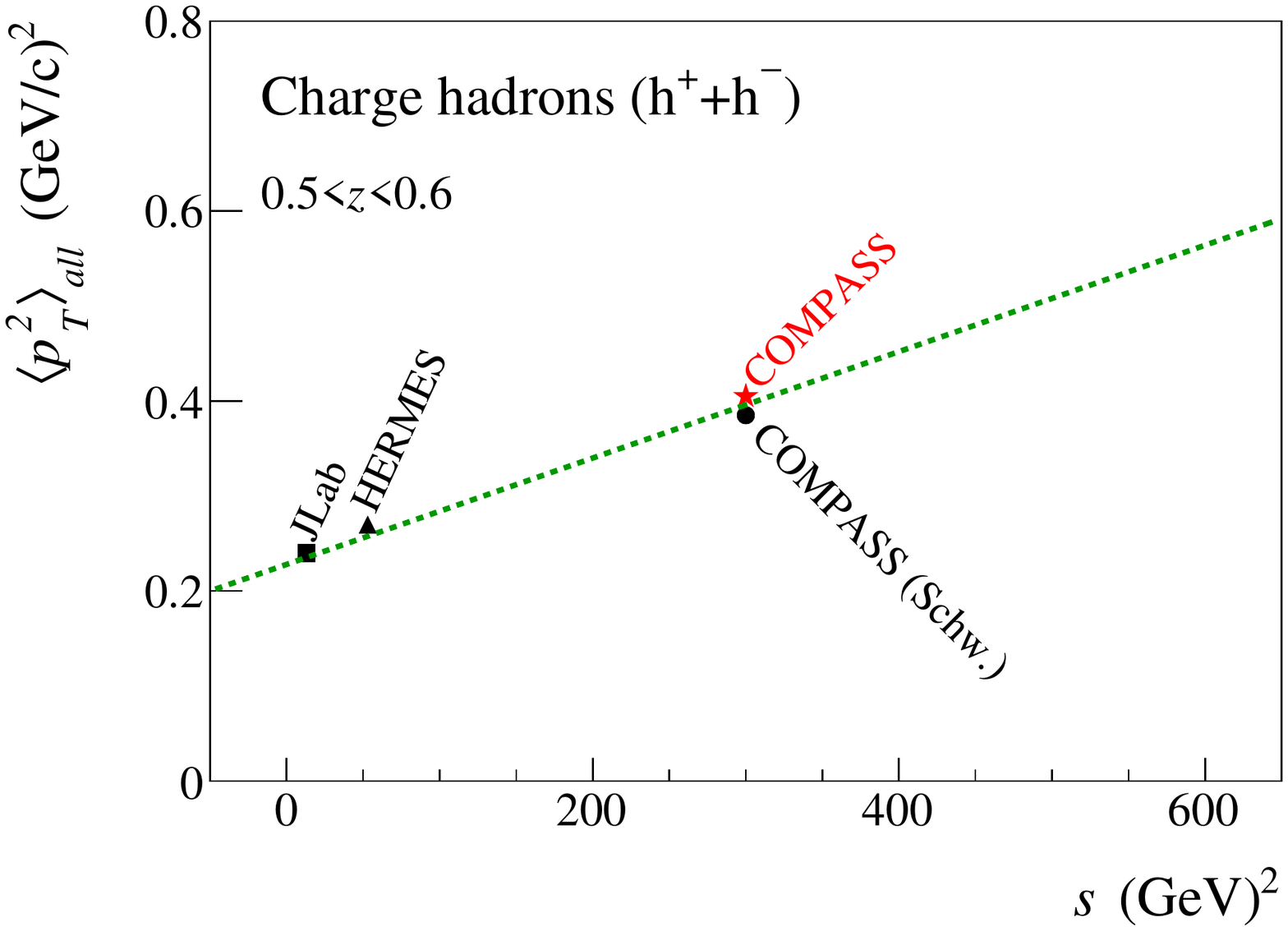}
 \end{center}
\caption{ The $p_T^2$ distribution of charged hadrons with $0.5<z<0.6$ used to
determine the acceptance corrected $\langle p_T^2 \rangle_{all}$ (left). The
$s$-dependence of $\langle p_T^2 \rangle_{all}$
from~Ref.\protect\cite{Schweitzer:2010tt} (right). The red star labeled COMPASS is
the value from this analysis, the black dot labeled COMPASS (Schw.) is the
value used in~Ref.\cite{Schweitzer:2010tt}, obtained from data not corrected for
acceptance.} 
\label{f_metz}
\end{figure}

It is interesting to compare the values and $W^2$-dependence of $\langle p_T^2
\rangle$ obtained from the fit at small $p_T$ with the values and
$W^2$-dependence of $\langle p_T^2 \rangle_{all}$. The $W^2$-dependence of
$\langle p_T^2 \rangle$, obtained from the fit in the bin $0.5 < z < 0.6$ is
shown in Fig.~\ref{f_pt2VSw2}, that one of $\langle p_T^2 \rangle_{all}$ in
Fig.~\ref{f_avptVSw2}. In addition to the data points, Fig.~\ref{f_avptVSw2}
shows lines, which represent fits of the data points assuming a linear function
of $\ln W^2$. Because of the $Q^2$-dependence, the last points are somewhat
below the fit. The authors of Ref.~\cite{Schweitzer:2010tt} first suggested that
$\langle p_T^2 \rangle_{all}$ should depend linearly on the $\mu N$ center of
mass energy squared $s$. They have verified their prediction with results from
three fixed target experiments: JLab, HERMES and COMPASS, see
Fig.~\ref{f_metz}. Fig.~\ref{f_metz}a shows the $ p_T^2 $ distribution of
charged hadrons with $0.5<z<0.6$ and integrated over $Q^2$ and $x_{Bj}$,
measured by COMPASS, which was used to determine the acceptance corrected
$\langle p_T^2 \rangle_{all}$. Fig.~\ref{f_metz}b taken from
Ref.~\cite{Schweitzer:2010tt} shows the dependence of $\langle p_T^2
\rangle_{all}$ on $s$. Their value for COMPASS, represented by the black dots,
was not corrected for acceptance. The new, acceptance corrected COMPASS value
$\langle p_T^2 \rangle_{all}$ added to Fig.~\ref{f_metz}b (red dot) is shown in
a recent paper~\cite{Pasquini:2011tk}, and used to quantify the $p_T$ broadening~\cite{collins85}
in a model to determine the Sivers and Boer-Mulders asymmetries at COMPASS and HERMES. The result of
the model of Pasquini and Schweitzer was closer to the COMPASS data when $p_T$
broadening is included. The authors of Ref.~\cite{Schweitzer:2010tt} also note
that $\langle p_T^2 \rangle_{all}$ may depend linearly on $W^2$ rather than
$s$. However, the dependence shown in Fig.~\ref{f_avptVSw2} is more compatible
with a linear dependence on $\ln W^2$ as was found by several experiments (see,
e.g.~\cite{Ashman:1991cj}). The relation is not well established and, as
mentioned inRef.~\cite{Schweitzer:2010tt}, the linear dependence on $s$ for
Drell-Yan which inspired their SIDIS prediction, could also be a linear
dependence on $\sqrt{s}$. Contrary to the case of $\langle p_T^2 \rangle_{all}$
in Fig.~\ref{f_avptVSw2}, the $W^2$-dependence of the fitted $\langle p_T^2
\rangle$ shown in Fig.~\ref{f_pt2VSw2} is much weaker, as expected, since
$\langle p_T^2 \rangle$ is assumed to be unaffected by pQCD, as opposed to
$\langle p_T^2 \rangle_{all}$.


\begin{figure}
 \begin{center}
  \includegraphics[width=0.48\textwidth]{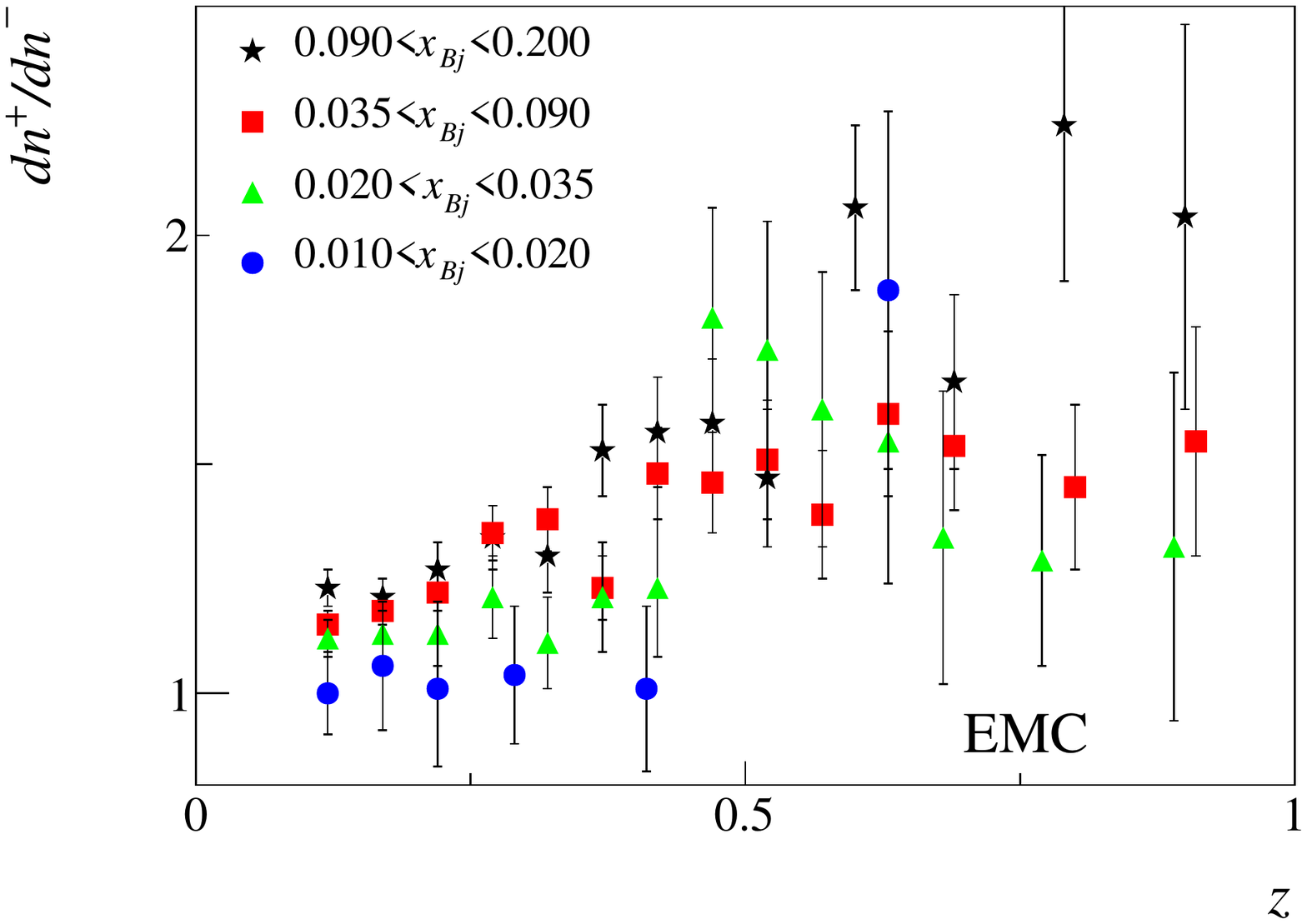} \hfill
  \includegraphics[width=0.48\textwidth]{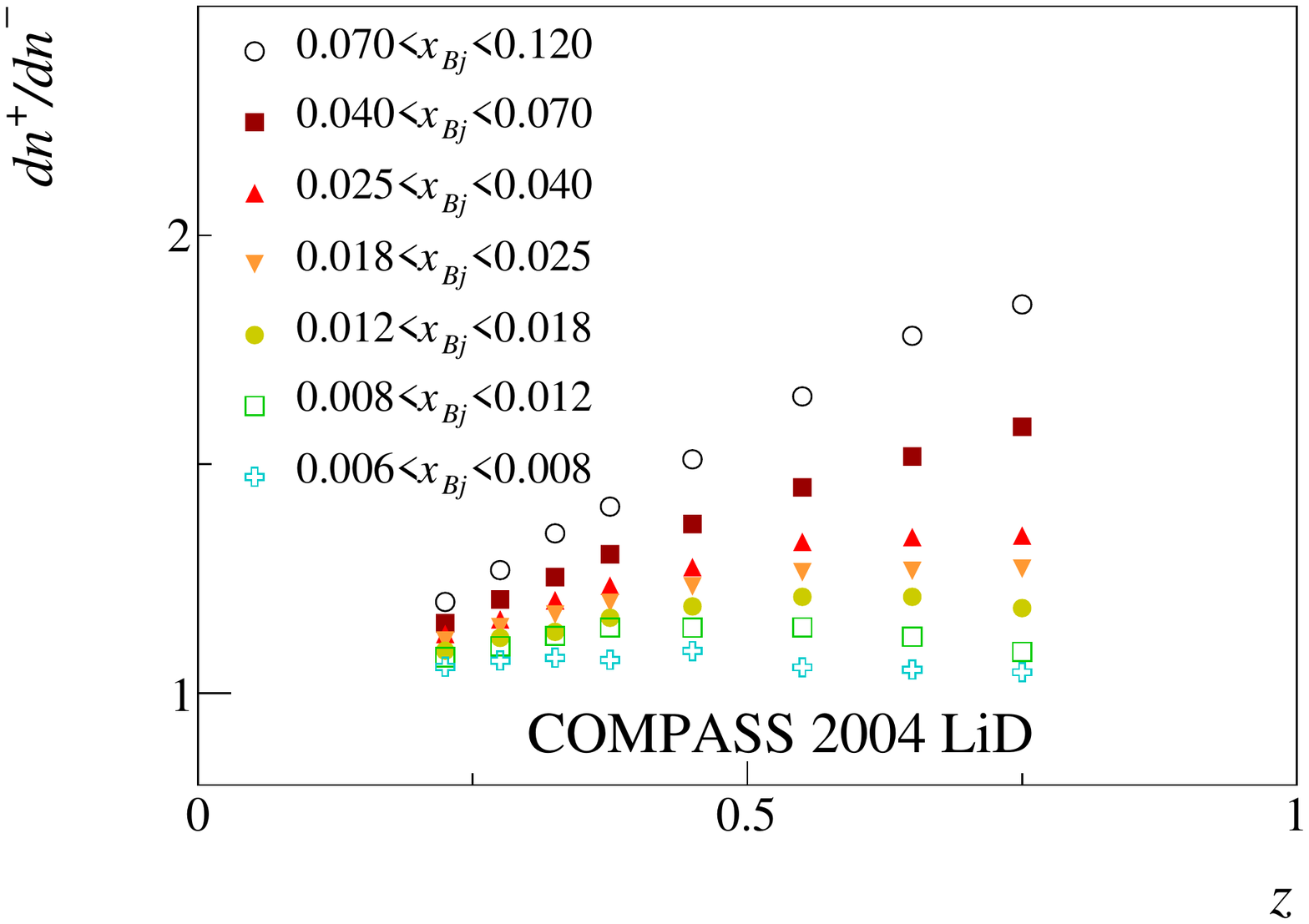}
 \end{center}
 \caption{Charged hadron multiplicity ratios $dn^{h+}/dn^{h-}$ as a function of
  $z$, for various $x_{Bj}$ bins, measured by EMC~\protect\cite{Ashman:1991cj}
  for $\mu \mathrm{D}$ (left) and COMPASS for $\mu ^6\mathrm{LiD}$ (right)
  interactions respectively. For COMPASS the results are based on part of the
  2004 collected statistics.}  
 \label{f_chgRatio}
\end{figure}

Another interesting observable is the ratio of the multiplicities of positive
and negative hadrons integrated over $p_T^2 $ and $Q^2$. The hadron
multiplicity ratios are shown in Fig.~\ref{f_chgRatio} and compared with
previous data taken by the EMC experiment~\cite{Ashman:1991cj}. COMPASS results
show clearly the $z$- and $x_{Bj}$-dependence, where the fraction of positive
hadrons increases with $x_{Bj}$ (getting closer to the valence region) and $z$
(more related to the energy of the struck parton). This behaviour can be qualitatively
connected with the fact that the positive valence quarks have a larger electric
charge than the negative ones.

The $z^2$-dependence of the fitted $\langle p_T^2 \rangle$ is of particular
interest. There are theoretical predictions allowing the extraction of the
intrinsic transverse momenta $k_{\perp}$ and $p_{\perp}$ from this
$z^2$-dependence~\cite{Anselmino:2005nn}.
At leading order QCD, assuming single photon exchange and an independent
fragmentation process, the hadron muoproduction cross section can be expressed
in terms of a hard muon-parton interaction cross section convoluted with two
unintegrated soft universal functions: $f_q(x_{Bj}, k_{\perp})$, the parton
distribution function of quark of flavor $q$ and $D^h_q(z, p_{\perp})$, the
fragmentation function defined as the number density of hadron $h$ resulting
from the fragmentation of a quark of flavor $q$. With the further assumption
that both $f_q(x_{Bj}, k_{\perp})$ and $D^h_q(z, p_{\perp})$ follow Gaussian
distributions with respect to the transverse momentum variables $k_{\perp}$ and
$p_{\perp}$, respectively, the cross section can be approximated~\cite{Anselmino:2005nn} at first order
in $\mathcal{O}(k_{\perp}/Q)$ by:

\begin{align}
\frac{d^4\sigma^{ \mu N \rightarrow \mu^{\prime}h X }}{dx_{Bj}dQ^2dzd{p}_{T}^2}
   \approx &\sum _q \frac{2\pi\alpha ^2e_q^2}{Q^4}
   f_q(x_{Bj}) D^h_q(z)\cdot
\left[1 + (1 - y)^2 \right]
   \frac{1}{\pi \langle p_T^2 \rangle_{q}}e^{-p_T^2/\langle p_T^2 \rangle_{q}},
\label{eq_xs_1srOrd_gauss}
\end{align}
where all the parameters describing the transverse momentum dependence of TMDs
for a given quark flavor $q$ are contained in $\langle p_T^2 \rangle_{q}$, through the relation:
\begin{equation}
\langle p_T^2 \rangle_{q} = \langle p_{\perp}^2 \rangle_{q} + z^2\langle
k_{\perp}^2 \rangle_{q}.
\label{eq_pt2VSz2}
\end{equation}
Here again, integration over the azimuthal angle has been performed. In
Ref.~\cite{Anselmino:2005nn} it was assumed that $\langle p_{\perp}^2 \rangle$ and $
\langle k_{\perp}^2 \rangle$ in Eq.~\ref{eq_pt2VSz2} are constants and
independent of the quark flavor. In general, they may both depend on $Q^2$ and
the active quark flavor $q$ while $\langle p_{\perp}^2 \rangle$ can depend
further on $z$ and the produced hadron type, and $ \langle k_{\perp}^2 \rangle$
may depend on $x_{Bj}$.

\begin{figure}
 \begin{center}
  \includegraphics[width=0.90\textwidth]{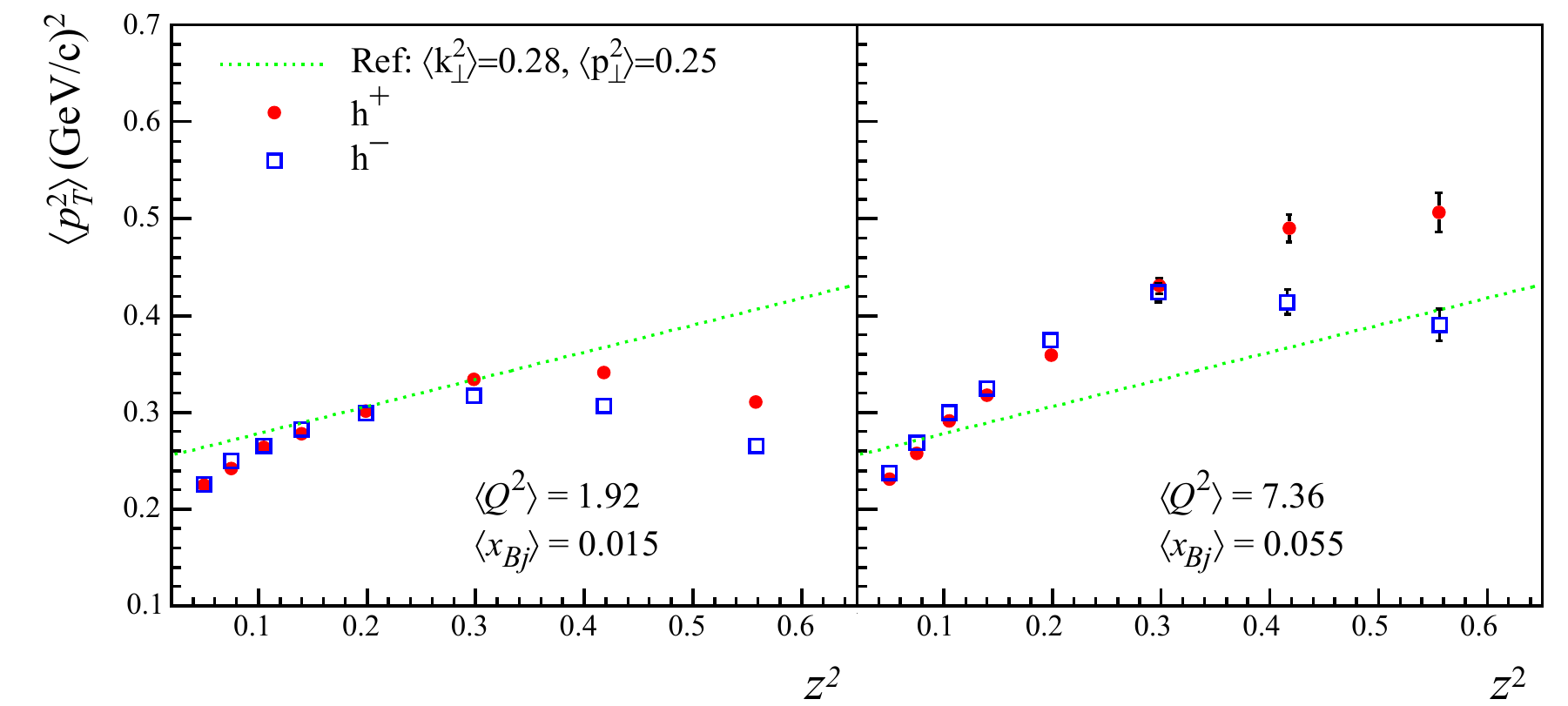}
 \end{center}
\caption{
$\langle p_T^2 \rangle$ vs $z^2$ for two ($Q^2$, $x_{Bj}$) intervals.
The corresponding average values $\langle Q^2\rangle$ (in units of \gomt\ and 
$\langle x_{Bj}\rangle$ are indicated in the figure.
The dotted green line corresponds to relation (\protect\ref{eq_pt2VSz2})
with constant $\langle k_{\perp}^2 \rangle$ and $\langle p_{\perp}^2 \rangle$
from Ref.~\protect\cite{Anselmino:2006rv}.
}
\label{f_pt2VSz2}
\end{figure}

\begin{figure}
 \begin{center}
  \includegraphics[width=0.90\textwidth]{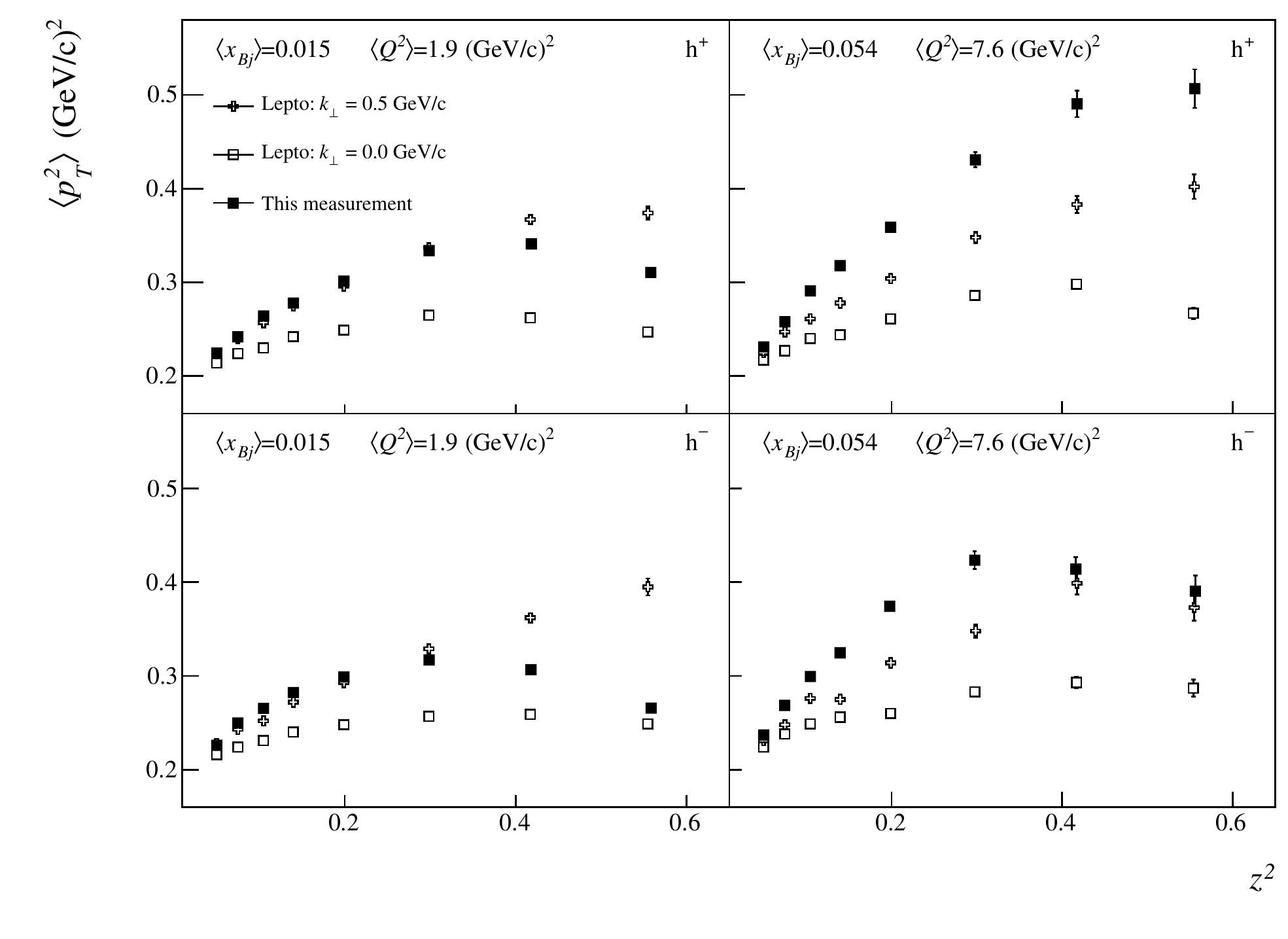}
 \end{center}
\caption{Comparison of the measured $\langle p_T^2 \rangle$ (full squares) with
a simulation using the MC event generator LEPTO for two bins of $Q^2$ and
$x_{Bj}$, for positive (top) and negative hadrons (bottom). Two cases were
simulated in the MC: Interactions without intrinsic transverse parton momenta
$\langle k_{\perp}^2 \rangle =0$ (open squares) and interactions with $\langle
k_{\perp}^2 \rangle = 0.25$ \gomt\ (open crosses). }
\label{PTdataVSMC}
\end{figure}

The observed dependence of the fitted $\langle p_T^2 \rangle$ on $z^2$ is shown
for two ($Q^2$, $x_{Bj}$) intervals in Fig.~\ref{f_pt2VSz2}. The relation
between $\langle p_T^2 \rangle$ and $z^2$ is certainly not linear as in
Eq.~\ref{eq_pt2VSz2}. It should be noted that the non linear behaviour of the 
$z^2$-dependence of $\langle p_T^2 \rangle$ was reproduced qualitatively in a
recent paper~\cite{Boglione:2011wm} by imposing kinematical constraints to the
model leading to Eq.~\ref{eq_xs_1srOrd_gauss}. A more general 
ansatz for the contributions of the intrinsic transverse momenta $ p_{\perp}$
and $ k_{\perp}$ to the measured hadron transverse momentum $ p_{T}$ is
\begin{equation}
\langle p_T^2 (z) \rangle = \langle p_{\perp}^2(z) \rangle + z^2\langle
k_{\perp}^2 \rangle ,
\label{eq_relation_pt_pperp_kperp_frag}
\end{equation}
where $\langle p_{\perp}^2(z) \rangle$ is a function of $z$ and should be taken
from other measurements. The dependence of $ k_{\perp}$ is still the same as in
Eq.~\ref{eq_relation_pt_pperp_kperp_frag}, with a constant average $\langle k_{\perp}^2 \rangle$. The
knowledge of $\langle p_{\perp}^2(z) \rangle$ could be taken from DIS event
generators which are supposed to incorporate all known properties of jet
fragmentation. In Fig.~\ref{PTdataVSMC} the measured values of $\langle p_T^2
\rangle$ are compared with those of a simulation using the event generator
LEPTO\footnote{For these simulations, MRST2004LO PDFs in LHAPDF 5.2.2 were used;
default LEPTO 6.5.1 and JETSET 7.4 settings, with the exception of LST(11)=122,
which includes target mass effects and the longitudinal structure function.}. 
Two cases were simulated in the MC: interactions without intrinsic
transverse parton momenta $\langle k_{\perp}^2 \rangle =0$ (open squares) and
interactions with $\langle k_{\perp}^2 \rangle = 0.25$ \gomt\ (open
crosses). For $\langle k_{\perp}^2 \rangle = 0.25$ \gomt, the agreement
between $\langle p_T^2 \rangle$ from simulated events and from data (full
squares) is striking for lower values of $Q^2$, apart from the highest
$z^2$ bins. For values of $Q^2$ larger than 4 \gomt, the data are
significantly above the simulation. The significant differences between
positive and negative hadrons at larger $z$ values are not reproduced by the MC
simulation. This comparison suggests that $\langle k_{\perp}^2 \rangle $ can be
extracted from the data, provided a detailed tuning of the jet fragmentation
parameters would be performed. In addition, it should be noted that the present
event selection includes all semi-inclusive events. Thus events which are not
due to DIS, i.e. the absorption of the virtual photon by a quark with subsequent
quark fragmentation, but to other mechanisms like diffractive vector meson
production are included in the sample. The treatment of this kind of
background to DIS requires special efforts, in order to extract $\langle
k_{\perp}^2 \rangle$. 

\section{Conclusion}
The main result of this study is the measurement of differential multiplicities
of charged hadrons produced in unpolarised SIDIS of muons off a $^6$LiD target.
The acceptance corrected multiplicities in a 4-dimensional ($z$, $p_T^2$,
$x_{Bj}$, $Q^2$) phase space are given in Ref.~\cite{HEPDATA} separately for
positively and negatively charged hadrons.

From these basic multiplicities more complex observables have been extracted.
The average $p_T^2$ over the entire $p_T$ range, $\langle p_T^2 \rangle_{all}$,
corrected for acceptance, has been provided for a comparison with other
experiments at different center of mass energy. The evolution of $\langle p_T^2
\rangle_{all}$ as a function of the invariant mass $W^2$ has been shown to
follow a linear dependence on $\ln W^2$ reasonably well. The ratio of positive
over negative hadrons is shown as a function of $z$ for bins in $x_{Bj}$ and
compared with previous EMC results.

The differential distributions at low $p_T^2$ have been fitted with an exponential
at different $z$ in order to obtain $\langle p_T^2 \rangle $. These data should
allow to determine the average intrinsic transverse momentum squared $\langle
k_{\perp}^2 \rangle$ in a framework based on QCD parton model and factorization.
The additional information needed to extract $\langle k_{\perp}^2 \rangle$ is
the intrinsic transverse momentum $\langle p_{\perp}^2 \rangle$ acquired during
fragmentation. This information may be derived from up-to-date Monte Carlo
simulations like LEPTO or PYTHIA, which incorporate all known properties of jet
fragmentation.

This analysis represents the first multidimensional study of hadron
multiplicities in unpolarised SIDIS at COMPASS. Further analyses, using the
COMPASS ability to identify hadrons, are under way to investigate the production
of pions and kaons.

\noindent \textbf{Acknowledgment}\\
\noindent We gratefully acknowledge the support of the CERN management and staff and the skill and effort of
the technicians of our collaborating institutes. Special thanks go to V. Anosov and V. Pesaro for their
technical support during the installation and the running of this experiment. This work was made possible
by the financial support of our funding agencies. 
We would like to thank Dr. Alessandro Bacchetta for his helpful comments.

\end{document}